\documentclass[10pt,twocolumn,cleanhead,cleanfoot]{asme2e}

\usepackage{empheq, amssymb}

\usepackage{lmodern}
\usepackage{amsmath}
\usepackage{amsfonts}
\usepackage{amssymb}
\usepackage{bm} 
\usepackage{txfonts}
\usepackage[T1]{fontenc}
\usepackage[utf8]{inputenc} 
\usepackage[dvipsnames]{xcolor} 
\usepackage[american]{babel} 
\usepackage{graphicx}
\usepackage{epstopdf}
\usepackage[noadjust]{cite} 
\usepackage{ltxcmds}
\usepackage{mathtools}
\usepackage{subcaption}
\usepackage{multirow}
\usepackage{blindtext}
\PassOptionsToPackage{hyphens}{url}
\usepackage{hyperref}
\usepackage{enumitem}
\usepackage{ifthen}
\usepackage{etoolbox}
\usepackage{cancel}
\usepackage{empheq}
\usepackage{fontawesome}
\usepackage{soul}
\usepackage{flushend} 
\usepackage{textcomp}
\usepackage{lipsum}

\definecolor{light-gray}{gray}{0.4}
\definecolor{box-gray}{gray}{1}

\hypersetup{
    unicode=false,          
    pdftoolbar=true,        
    pdfmenubar=true,        
    pdffitwindow=false,     
    pdfstartview={FitV},    
    pdftitle={A case study comparing both stochastic and worst-case robust control co-design under different control structures},    
    pdfauthor={Saeed Azad and Daniel R. Herber},     
    pdfnewwindow=true,      
    colorlinks=true,
    allcolors=blue
}

\usepackage{nomencl}

\renewcommand\nomgroup[1]{%
  \item[\bfseries
  \ifstrequal{#1}{V}{ Select Variables}{%
  \ifstrequal{#1}{B}{ Subscripts}{%
  \ifstrequal{#1}{P}{ Notation}{%
  \ifstrequal{#1}{A}{ Acronyms}{}}}}]
}

\makenomenclature



\usepackage{dblfloatfix}
\usepackage{float}

\definecolor{block-gray}{gray}{0.95}

\usepackage[framemethod=TikZ]{mdframed}

\usepackage{xpatch}

\makeatletter
\xpatchcmd{\endmdframed}
  {\aftergroup\endmdf@trivlist\color@endgroup}
  {\endmdf@trivlist\color@endgroup\@doendpe}
  {}{}
\makeatother

\usepackage{fontawesome}
\usepackage{titlesec}

\usepackage{hyperref}

\newcommand{\doi}[1]{{doi:~\href{http://doi.org/#1}{#1}}\rmFullStop}

\newcommand*{\rmFullStop}{\rmifnextchar{.}{}{}}

\makeatletter
\newcommand{\rmifnextchar}[3]{%
  \begingroup
  \ltx@LocToksA{\endgroup#2}%
  \ltx@LocToksB{\endgroup#3}%
  \ltx@ifnextchar{#1}{%
    \def\next{\the\ltx@LocToksA}%
    \afterassignment\next
    \let\scratch= %
  }{%
    \the\ltx@LocToksB
  }%
}
\makeatother

\usepackage{colortbl}
\definecolor{light-gray}{gray}{0.6}

\newcommand{\xsection}[1]{\section[#1]{\MakeUppercase{#1}}}

\usepackage{accents}

\definecolor{needcolor}{HTML}{C62828}

\usepackage[most]{tcolorbox} 
\definecolor{block-gray}{gray}{0.95}

\AtBeginDocument{%
 \abovedisplayskip=4pt plus 1pt minus 1pt
 \abovedisplayshortskip=0pt plus 3pt
 \belowdisplayskip=4pt plus 1pt minus 1pt
 \belowdisplayshortskip=0pt plus 3pt
}

\newtcolorbox{xreviewer}{%
	empty,
    borderline west = {4pt}{0pt}{gray},
    boxrule = 0pt,
    boxsep = 0pt,
    breakable,
    colback = block-gray,
    enhanced,
    frame hidden,
    left skip = 0pt,
    notitle,
    parbox = false,
    sharp corners,
}
\newtcolorbox{xresponse}{%
	empty,
    boxsep = 0pt,
    breakable,
    frame hidden,
    notitle,
    parbox = false,
}
\newtcolorbox{xchange}{%
    boxrule = 1pt,
    boxsep = 0pt,
    breakable,
    colframe = black,
    enhanced jigsaw,
    interior hidden,
    notitle,
    parbox = false,
    arc=0pt,
    outer arc=0pt,
    after skip=20pt plus 2pt,
}

\usepackage{catchfilebetweentags}
\makeatletter
\def\CatchFBT@Fin@l#1[#2]{%
   \begingroup
      \makeatletter #2%
      \scantokens\expandafter{%
         \expandafter\CatchFBT@tok\expandafter{\the\CatchFBT@tok}}%
      \CatchFBT@IsAToken{#1}
         {\global#1\expandafter{\the\CatchFBT@tok}}
         {\xdef#1{\the\CatchFBT@tok}}%
      \ifx\CatchFBT@tok#1\else\global\CatchFBT@tok{}\fi
   \endgroup
}
\makeatother

\usepackage{xcolor}
\newcommand{\parm}{\mathord{\color{black!33}\bullet}}%

\usepackage{nomencl}
\makenomenclature
\usepackage{mathrsfs}
\usepackage{euscript}
\usepackage{mathtools}

\usepackage{accents}

\usepackage{units}
\usepackage{xspace}
\usepackage{tablefootnote}
\allowdisplaybreaks
\usepackage{caption}
\usepackage{subcaption}
\usepackage{tablefootnote}

\usepackage[normalem]{ulem}

\definecolor{MyFavoriteBlue}{HTML}{4E79A6}

\newcommand{\ti}{} 

\let\svthefootnote\thefootnote

\newcommand{\pd}{\bm{d}}      

\title{A case study comparing both stochastic and worst-case robust control co-design under different control structures} 

\author{Saeed~Azad\thanks{Corresponding author} 
\affiliation{
Postdoctoral Fellow\\
Department of Systems Engineering \\
Colorado State University\\
Fort Collins, CO 80523 \\
Email:~\texttt{\href{mailto:saeed.azad@colostate.edu}{saeed.azad@colostate.edu}}\\ 
}
}

\author{Daniel~R.~Herber
\affiliation{
Assistant Professor\\
Department of Systems Engineering \\
Colorado State University \\
Fort Collins, CO 80523 \\
Email:~\texttt{\href{mailto:daniel.herber@colostate.edu}{daniel.herber@colostate.edu}}
}
}

\usepackage{amsmath}

\begin{document}
 \setlength{\parskip}{0pt}
 \setlength{\parsep}{0pt}
 \setlength{\headsep}{0pt}
\setlength{\topsep}{0pt}

\abovedisplayshortskip=3pt
\belowdisplayshortskip=3pt
\abovedisplayskip=3pt
\belowdisplayskip=3pt

\titlespacing*{\section}{0pt}{18pt plus 1pt minus 1pt}{3pt plus 0.5pt minus 0.5pt}

\titlespacing*{\subsection}{0pt}{9pt plus 1pt minus 0.5pt}{1pt plus 0.5pt minus 0.5pt}

\titlespacing*{\subsubsection}{0pt}{9pt plus 1pt minus 0.5pt}{1pt plus 0.5pt minus 0.5pt}

\maketitle

\let\thefootnote\relax\footnote{The original version of this paper was presented in ASME IMECE in 2022 (IMECE2022-95229). This submission entails significant changes listed in the cover letter.}\addtocounter{footnote}{-1}\let\thefootnote\svthefootnote

\begin{abstract}\noindent
\textit{As uncertainty considerations become increasingly important aspects of concurrent plant and control optimization, it is imperative to identify and compare the impact of uncertain control co-design (UCCD) formulations on their associated solutions.
While previous work has developed the theory for various UCCD formulations, their implementation, along with an in-depth discussion of the structure of UCCD problems, implicit assumptions, method-dependent considerations, and practical insights, is currently missing from the literature.
Therefore, in this study, we address some of these limitations by focusing on some UCCD formulations, with an emphasis on optimal control structures, and uncertainty propagation techniques. 
Specifically, we propose three optimal control structures for UCCD problems: (i)
open-loop multiple-control (OLMC), (ii) multi-stage control (MSC), and (iii) 
open-loop single-control (OLSC).
Stochastic in expectation UCCD (SE-UCCD) and worst-case robust UCCD (WCR-UCCD) formulations, which are motivated by probabilistic and crisp representations of uncertainties, respectively, are implemented for a simplified strain-actuated solar array case study.
Solutions to the OLMC SE-UCCD problem are obtained using two uncertainty propagation techniques: generalized Polynomial Chaos expansion (gPC) and Monte Carlo simulation (MCS).
The OLMC and MSC WCR-UCCD problems are solved by leveraging the structure of the linear program, leading to polytopic uncertainties.
To highlight the importance of uncertainty considerations in early-stage design, the closed-loop reference-tracking response of resulting systems is also investigated. 
Insights gained from such studies underscore the role of the control structure in managing the trade-offs between risk and performance, as well as meeting problem requirements.
The results also emphasize the benefits of efficient uncertainty propagation techniques, such as gPC, for dynamic optimization problems.     
}
\end{abstract}

\vspace{1ex}
\noindent \textit{Keywords:~uncertain control co-design; control structures in UCCD; stochastic UCCD; worst-case robust UCCD; multi-stage MPC in UCCD}

\xsection{Introduction}\label{sec:introduction}

In uncertain control co-design (UCCD) problems, uncertainty is a challenging and unavoidable aspect of modeling system behavior and realizing pragmatic design solutions.
Previous work has offered a broad overview of uncertainties, their interpretation, and ways to integrate them into various UCCD formulations~\cite{Azad2022}.
This overview resulted in several UCCD problems motivated by concepts from stochastic programming~\cite{ruszczynski2003stochastic, powell2019unified}, robust optimization~\cite{beyer2007robust, gorissen2015practical, bertsimas2011theory}, and fuzzy programming~\cite{zadeh1996fuzzy, zadeh1978fuzzy, liu2009theory}.
However, effective solution strategies are required to answer the design optimization question put forth by a novel UCCD formulation.
In this article, we start to fill this gap by focusing on implementations for two unique and illustrative UCCD formulations with a focus on open-loop control structures and uncertainty propagation techniques.

We first introduce and discuss the optimal control structures under uncertainties.
In this context, we use the term \textit{control structure} to refer to various ways in which control trajectories can be structured in the UCCD problem in a meaningful manner (and is different from CCD coordination strategies, such as simultaneous and nested~\cite{herber2019nested, allison2014special, sundarrajan2021towards}).
The need for the discussion on control structures is motivated by the role that optimal control trajectories play in UCCD problems.
Although such control structures are commonly used in different control engineering applications, their role in managing uncertainties (i.e.,~risk versus performance), and their impact on the integrated robust solution in the UCCD framework are not well discussed, nor clearly demonstrated \textemdash{} particularly for boundary-value UCCD problems.

In this article, we introduce three distinct control structures that highlight the role of control decisions in response to uncertainties: (i) \textit{open-loop multiple-control} (OLMC), (ii) \textit{multi-stage control} (MSC), and (iii) \textit{open-loop single-control} (OLSC).
These structures are related to concepts from stochastic control~\cite{powell2019unified, nakka2021trajectory, berning2018rapid}, robust multi-stage model predictive control (MPC) \cite{lucia2017rapid,lucia2020stability},
and worst-case robust control~\cite{diehl2008numerical, diehl2006approximation, ma2001worst, nash2021robust}, respectively.  
Insights from our discussion on control structures, presented in Sec.~\ref{sec:section2}, are expected to provide a more meaningful approach toward meeting problem requirements in the presence of uncertainties.

In the next step, a simple CCD problem is used as a case study (adopted from Refs.~\cite{herber2017unified, chilan2017co}) and modified to include all time-independent, natural uncertainties. 
These uncertainties stem from plant optimization variables, initial states, and problem data and result in a UCCD problem that can be formulated like any of the forms discussed in Ref.~\cite{Azad2022}.
The simple strain-actuated solar array (SASA) UCCD problem~\cite{herber2017unified, chilan2017co} was selected due to its simplicity in both implementation and the interpretation of results.
In this study, we develop and implement the simple SASA problem for two different UCCD formulations: stochastic in expectation UCCD (SE-UCCD) and worst-case robust UCCD (WCR-UCCD).

Since the solution of UCCD problems with probabilistic uncertainties requires an appropriate uncertainty propagation (UP) method, we also introduce Monte Carlo Simulation (MCS) and the non-intrusive, collocation-type, generalized Polynomial Chaos (gPC).
For problems with a crisp (also known as bounded) representation of uncertainties, we use polytopic uncertainties and their implications on linear programs.
The role of feedback information in the presence of crisp uncertainties is investigated through the MSC structure.   
Finally, open-loop solutions from these implementations are utilized to synthesize a closed-loop feedback controller.
The analysis of closed-loop simple SASA systems in the presence of uncertainties points to critical distinctions in each system's behavior \textemdash{} emphasizing the impact of early-stage uncertainty considerations in the design process.
By avoiding problem-dependent complexities, this article allows for a straightforward comparison of such formulations \textemdash{} highlighting the impact of problem formulation and method-dependent decisions on the UCCD solution.

\subsection{Deterministic Control Co-design}
\label{subsec:DCCD}
The nominal continuous-time, deterministic, all-at-once (AAO), simultaneous, CCD problem is formulated as \cite{herber2019nested, herber2020uses}:
\begin{subequations}
\label{Eqn:DCCD}
\begin{align}
\underset{\bm{u}, \bm{\xi}, \bm{p}}{\textrm{minimize:}} \quad & o=\displaystyle\int_{t_0}^{t_f}  \ell(t,\bm{u},\bm{\xi}, \bm{p}, \pd)\,\mathrm{d}t + m (\bm{p}, \bm{\xi}_{0},\bm{\xi}_{f}, \pd)\label{Eqn:DCCD_obj}\\
\textrm{subject to:} 
\quad & \bm{g}(t,\bm{u},\bm{\xi}, \bm{p}, \bm{\xi}_{0}, \bm{\xi}_{f},\pd)\leq \bm{0}\label{Eqn:DCCD_ineq}\\
\quad & \bm{h}(t,\bm{u},\bm{\xi}, \bm{p}, \bm{\xi}_{0}, \bm{\xi}_{f}, \pd) = \bm{0} \label{Eqn:DCCD_eq}  \\
\quad & \dot{\bm{\xi}} - \bm{f}(t,\bm{u},\bm{\xi}, \bm{p},\bm{\xi}_{0}, \bm{\xi}_{f}, \pd) = \bm{0} \label{Eqn:DCCD_dyn}\\ 
\textrm{where:}
\quad & \bm{\xi}(t_0) = \bm{\xi}_{0},~ \bm{\xi}(t_f) = \bm{\xi}_{f},~ \bm{u}(t) = \bm{u},~ \bm{\xi}(t) = \bm{\xi} \label{Eqn:DCCD_ini} \\
& \pd(t) = \pd \nonumber
\end{align}
\end{subequations}

\noindent where $t \in [t_0, t_f]$ is the time horizon, $\{ \bm{u}, \bm{\xi}, \bm{p} \}$ are the collection of optimization variables including the open-loop control trajectories $\bm{u}(t) \in \mathbb{R}^{n_u}$, state trajectories $\bm{\xi}(t) \in \mathbb{R}^{n_s}$, and the vector of time-independent optimization variables $\bm{p} \in \mathbb{R}^{n_p}$, respectively.
Note that $\bm{p}$ may entail plant optimization variables $\bm{p}_{p}$ and/or time-independent control optimization variables $\bm{p}_{c}$~\cite{fathy2003nested, herber2017unified} (i.e.,~gains), such that $\bm{p} = [\bm{p}_{p}, \bm{p}_c]$.
The objective function $o(\cdot)$ is composed of the Lagrange term $\ell(\cdot)$ and the Mayer term $m(\cdot)$.
The vectors of inequality and equality constraints are described by $\bm{g}(\cdot)$ and $\bm{h}(\cdot)$, respectively.
The transition or state derivative function $\bm{f}(\cdot)$ describes the evolution of the system through time in terms of a set of ordinary differential equations (ODEs). 
All of the time-dependent or time-independent data associated with the problem formulation, such as problem constants, environmental signals, initial/final times, etc., is represented through $\pd \in \mathbb{R}^{n_{d}}$.

In the remainder of this article, we assume that constraints associated with the initial and final conditions $\{\bm{\xi}_{0}, \bm{\xi}_{f}\}$ are already included in $\bm{h}(\cdot)$ or $\bm{g}(\cdot)$. 
We will often drop the explicit dependence on $t$ from time-dependent quantities such as control and state trajectories, as well as the problem data.
For more details on deterministic CCD, the readers are referred to Refs.~\cite{allison2014special, herber2019nested}.

\subsection{Uncertain CCD Problem Formulation}
\label{subsec:UCCCformulation}

Using $\tilde{\parm}$ to distinguish uncertain from deterministic variables, we introduce an AAO, continuous-time, simultaneous UCCD formulation in the probability space \cite{Azad2022}: 
\begin{subequations}
 \label{Eqn:UCCD}
 \begin{align}
 \underset{\tilde{\bm{u}}, \tilde{\bm{\xi}}, \tilde{\bm{p}}}{\textrm{minimize:}}
 \quad & \mathbb{E}\left [\bar{o} ( t, \tilde{\bm{u}}, \tilde{\bm{\xi}}, \tilde{\bm{p}}, \tilde{\pd}) \right]  \label{UCCD_obj} \\
 \textrm{subject to:} \quad
& \mathbb{E}\left [ \bar{\bm{g}} ( t, \tilde{\bm{u}}, \tilde{\bm{\xi}}, \tilde{\bm{p}}, \tilde{\pd}) \right] \leq {\bm{0}} \label{UCCD_ineq} \\
& {\bm{h}}(t, \tilde{\bm{u}}, \tilde{\bm{\xi}}, \tilde{\bm{p}}, \tilde{\pd})={\bm{0}}  \label{UCCD_eq}  \\
& \dot{\tilde{{\bm{\xi}}}}(t) - {\bm{f}}(t, \tilde{\bm{u}}, \tilde{\bm{\xi}}, \tilde{\bm{p}}, \tilde{\pd}) = \bm{0} \label{UCCD_dyn} \\
\textrm{where:} \quad & \tilde{\bm{u}} = \tilde{\bm{u}}(t) \in \mathcal{U}_{t},~\tilde{\bm{\xi}} = \tilde{\bm{\xi}}(t) \in \mathcal{U}_{t} \label{UCCD_where}\\
& \tilde{\bm{p}} \in \mathcal{U}_{\ti},~\tilde{\pd} = \tilde{\pd}(t) \in \mathcal{U}_{t} \notag 
\end{align} 
\end{subequations}
\noindent
where the expectation of $\bar{{o}}(\cdot)$, which is a function composition of $o(\cdot)$, is optimized over the set of optimization variables $(\tilde{\bm{u}}, \tilde{\bm{\xi}}, \tilde{\bm{p}})$, and is subject to the expectation of $\bar{\bm{g}}(\cdot)$ (i.e.,~a function composition of $\bm{g}(\cdot)$), analysis-type equality constraints ${\bm{h}}(\cdot)$, and uncertain dynamic system equality constraints in Eq.~(\ref{UCCD_dyn}). 
Note that $\mathbb{E}[\bar{o}(\cdot)]$ and $\mathbb{E}[\bar{\bm{g}}(\cdot)]$ may be any of the variations that are discussed in Ref.~\cite{Azad2022} (such as the nominal, worst-case, expected value, etc.).
This formulation includes the vector of uncertain control processes
$\tilde{\bm{u}}(t) \in  \mathcal{U}_{t}$, uncertain state processes $\tilde{\bm{\xi}}(t) \in \mathcal{U}_{t}$, time-independent uncertain optimization variables $\tilde{\bm{p}} \in \mathcal{U}_{\ti}$, and
time-dependent $\tilde{\pd}(t) \in \mathcal{U}_{t}$ and/or time-independent uncertain problem data $\tilde{\pd} \in \mathcal{U}_{\ti}$.
Here, $\mathcal{U}_{\ti}$ and $\mathcal{U}_{t}$ refer to time-independent and time-dependent uncertainty sets.

Through the appropriate selection of the objective function and constraints, specialized formulations such as stochastic in expectation (SE-UCCD), stochastic chance-constrained (SCC-UCCD), worst-case robust (WCR-UCCD), probabilistic robust (PR-UCCD), fuzzy expected value (FE-UCCD), and possibilistic chance-constrained (PCC-UCCD) can be derived. All of these formulations are discussed in detail in Ref.~\cite{Azad2022}. 

The remainder of this article is organized in the following manner: 
Sec.~\ref{sec:section2} describes some considerations in implementing UCCD problems; 
UP methods such as MCS and gPC are introduced in Sec.~\ref{sec:section3}; 
SE-UCCD and WCR-UCCD formulations associated with the simple SASA case study are described in Sec.~\ref{sec:section4};
and Sec.~\ref{sec:section5} presents results and discussion from such implementations.
Finally, Sec.~\ref{sec:conclusion} presents the conclusions.

\xsection{UCCD Implementation }
\label{sec:section2}

In this section, we start by discussing the optimal control structures under uncertainties in UCCD problems.
These structures attempt to answer inherently different questions and have been used in separate studies in the literature \cite{azad2020single, azad2021robust, fisher2011optimal, boutselis2016stochastic, cottrill2012hybrid, nagy2004open, meyers2019koopman, gonzalez2016wind, nash2021robust, behtash2021reliability}.
While these structures have a significant impact on the problem implementation and its associated solution, they have not been collectively discussed in the literature to the best of the authors' knowledge.
This section also highlights some subtle implementation considerations and extends the discussion to the effectiveness of various coordination strategies, such as simultaneous, nested, and direct single shooting (DSS), for general UCCD problems \cite{herber2019nested, allison2014special, sundarrajan2021towards, rao2009survey}. 

\subsection{Control Structures Under Uncertainties}
  
To respond to the presence of uncertainties and manage their impact on a UCCD problem, various control structures can be utilized.
Consider an uncertain dynamic system 
characterized through a set of dynamical systems belonging to the set $\mathcal{P}_{d}$.
This representation may be accomplished in various ways, including an appropriate uncertainty propagation technique 
Depending on the choice of the robust horizon parameter $\Delta t_{r}$, the control structure for the specific CCD problem in $t\in[t_{0}, t_{f}]$ can be formulated in continuous-time through the following conditions:

\begin{equation}
    \begin{aligned}
    \label{Eq:CS}
             &\bm{u}^{i}(t) \coloneqq
                \begin{cases}
                    \bm{u}^{j}(t) \quad & t < t_{0} + \Delta t_{r}\\
                    \text{admissible }({\mathcal{U}_{a}}) \quad & t \geq t_{0}+\Delta t_{r}\\
                    \text{admissible }({\mathcal{U}_{a}}) \quad & \Delta t_{r}~\text{is undefined}\ 
                \end{cases}
             & \forall~ i,j \in \mathcal{P}_{d} 
    \end{aligned}
\end{equation}

\begin{figure}[t]
\centering
\includegraphics[width=\columnwidth]{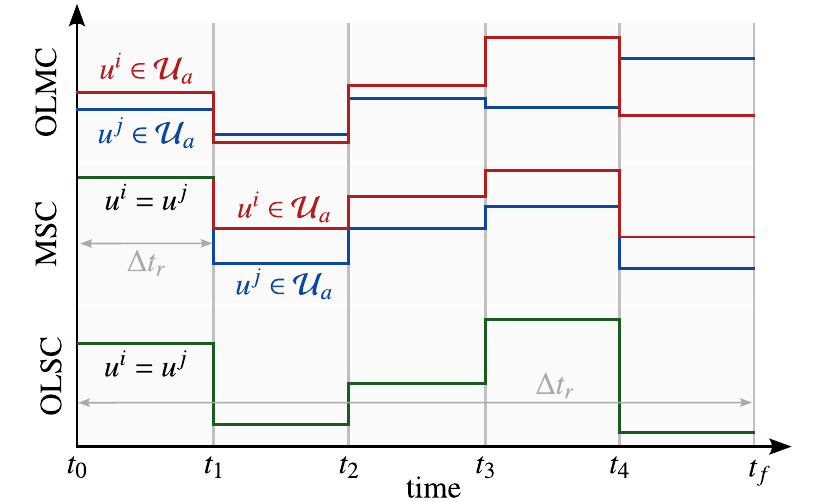}
\caption{Comparison of OLSC, MSC, and OLMC structures with $\mathcal{U}_{a}$ describing the admissible control set, and $\Delta t_{r}$ being the robust horizon.}
\label{fig:CS}
\end{figure}

\noindent
Now, for any choice of $\Delta t_{r} \in [0, t_{f}-t_{0}]$, the control structure requires a single control trajectory for all uncertain plants throughout the robust horizon $[t_{0}, t_{0}+\Delta t_{r})$, but allows for a distinct optimal trajectory for each uncertainty realization throughout the rest of the time horizon $[t_{0}+\Delta t_{r}, t_{f}]$.
This control structure is motivated by the robust multi-stage approach that is often used in MPC.
In the limits of this formulation, we can find two specific cases: 
(i) no robust horizon is specified (i.e.~$\Delta t_{r} \coloneqq \text{undefined}$). In this case, a distinct open-loop optimal trajectory is required for each uncertainty realization throughout the entire horizon.
We refer to this as the \textit{open-loop multiple control structure (OLMC)}. 
Alternatively, (ii) the robust horizon can be specified as the entire time horizon.
This control structure requires a single open-loop control trajectory for all uncertain systems throughout the entire time horizon $[t_{0}, t_{f}]$, and we refer to it as the \textit{open-loop single control (OLSC)} structure.
An illustrative comparison of OLMC, MSC, and OLSC is shown in Fig.~\ref{fig:CS}.
In the following sections, we discuss these structures and their interpretations in more detail.

\subsubsection{Open-Loop Multiple-Control (OLMC).}~\label{subsec:OLMC}
OLMC is based on the idea that every realization of uncertainty should elicit a distinct optimal control response from the UCCD problem and is implemented in Refs.~\cite{cottrill2012hybrid, kitapbayev2015stochastic, chai2019solving, meyers2019koopman, cottrill2011hybrid, matsuno20134d}.
Even when control trajectories are not the source of uncertainty or when the control loop is not closed, this approach results in an aggregate of optimal control solutions that provide additional insights into the limits of the system's performance. 
As an example, OLMC can be used to discover the upper system performance limits at the early-stage design of a vehicle active-suspension system, such as the maximum actuator force, maximum sprung mass acceleration, etc. 
Therefore, the collection of these control trajectories can be used to provide insights into defining the control architecture.

An important feature of OLMC is that, compared to OLSC, it enables a less conservative UCCD formulation by allowing the designer to exploit the control decision space more freely in response to uncertainties.
One way to view OLMC is to consider the closed-form solution to the optimal supervisory control for an arbitrary UCCD problem.
Since the solution can be directly written as a function of other problem elements, such as plant optimization variables or problem data, then uncertainties in these quantities elicit a range of optimal control responses.
As an example, consider a system in which uncertainties stem from plant optimization variables. 
While it is possible to change the plant design in response to uncertainties in order to achieve reliability or robustness (using methods based on reliability-based CCD \cite{azad2020single, cui2020comparative} or robust CCD \cite{azad2020robust}, respectively), it might be more cost-effective to leverage the control effort, including its limits, to achieve such criteria.
Therefore, OLMC may be more suitable for early-stage design, where plant and control spaces are explored for performance optimality, as well as reliability or robustness. 
 
The OLMC idea introduced here has some key differences from various control trajectories generated in an arbitrary closed-loop controller.
When the control loop is closed, different realizations of uncertainties result in different trajectories (which is also the case with OLMC). 
However, not all of these trajectories result in optimal behavior.
This outcome is in contrast with the OLMC structure, which seeks to generate control trajectories that are optimal for every realization of the uncertainties.  
Formally connecting these OLMC trajectories to desired controller structures (e.g., closed-loop tracking controllers) is a critical task for realization.

The implementation of OLMC structure in the simultaneous coordination strategy for any general UCCD problem with at least plant uncertainty\footnote{With plant uncertainty, the UP method needs to be constructed inside the optimization problem and evaluated at every iteration of the optimization solver. This results in a problem structure that cannot be solved sequentially through a simultaneous coordination strategy.\vspace{2em}} requires expanding the number of control variables to match the number of samples or function evaluation points.
This can result in a prohibitively large UCCD problem.
However, using a nested coordination strategy, the inner-loop optimal control subproblems associated with sample points become completely independent and can be solved more efficiently.
In other words, control variables, along with their associated dynamic equations, are decoupled and can be solved in parallel within the nested coordination strategy.
For this reason, all of the instances of OLMC-UCCD implementations in this study use a nested coordination strategy.

\subsubsection{Multi-Stage Control (MSC).}~\label{subsec:RMPC}
MSC is a robust control structure that defines a prediction and robust horizon to characterize the evolution of the uncertain system through a scenario tree \cite{lucia2017rapid}.
Uncertainties result in the creation of new branches in the scenario tree nodes for any chosen control input.
The robust horizon allows for the propagation of branches in (discrete) time to further account for both uncertainties and (future) control inputs.      
Through this structure, one can compute the (here and now) optimal values of the first-stage control variables by accounting for any potential adaptations of future inputs (in the robust horizon) through feedback.

Since the first-stage optimal control decisions are to be implemented for the actual system, they must be similar among all scenarios. 
However, future control decisions within any of the branches can vary in response to uncertainties.  
Due to the presence of feedback, the resulting solution from MSC is less conservative than the open-loop worst-case robust UCCD and the open-loop min-max nonlinear model predictive control \cite{lucia2017rapid}.

In terms of implementation, MSC requires solving a sequence of moving horizon (or receding horizon) optimal control problems for every candidate plant. 
Therefore, the nested coordination strategy is especially suitable for this structure.
In the presence of terminal conditions, the sequence of inner-loop optimal control problems are formulated as receding horizon problems.

\subsubsection{Open-Loop Single-Control (OLSC).}~ \label{subsec:OLSC}
In this structuring, the goal is to find a single control command that is optimal with respect to some criteria, such as the expectation of the objective function or the worst-case uncertainty realization.
In that sense, OLSC is closely related to concepts from robust control theory where the single robust control command is designed to perform well under a range of circumstances. 

For example, OLSC can generate an open-loop optimal control trajectory for robot motion planning, which is then followed using a tracking controller. 
While OLSC is particularly suitable for generating open-loop trajectories for use in trajectory-tracking applications, this structure has some inherent limitations.

For instance, consider a general UCCD problem with initial and terminal boundary conditions. 
This class of problems, which is referred to as boundary-valued UCCD, has an extensive application in various fields of engineering.
As an example, a boundary-valued UCCD problem might attempt to bring a vehicle to rest, an aircraft back to an airfield, the energy of a system to zero, or the temperature in a heat-transfer system to a specific value \textemdash{} all in the presence of uncertainties.
For simplicity, but without any loss of generality, assume that uncertainties in our boundary-valued UCCD problem originate only from uncertain plant optimization variables.
In this case, it becomes immediately evident that the OLSC structure cannot satisfy all of the prescribed initial and terminal boundary conditions in the presence of uncertainties. 
In other words, there's no single control that can satisfy the prescribed boundary conditions for all of the realizations of plant uncertainty because the OLSC problem is \textit{over-constrained}. 

\begin{figure}[t]
\centering
\includegraphics[width=\columnwidth]{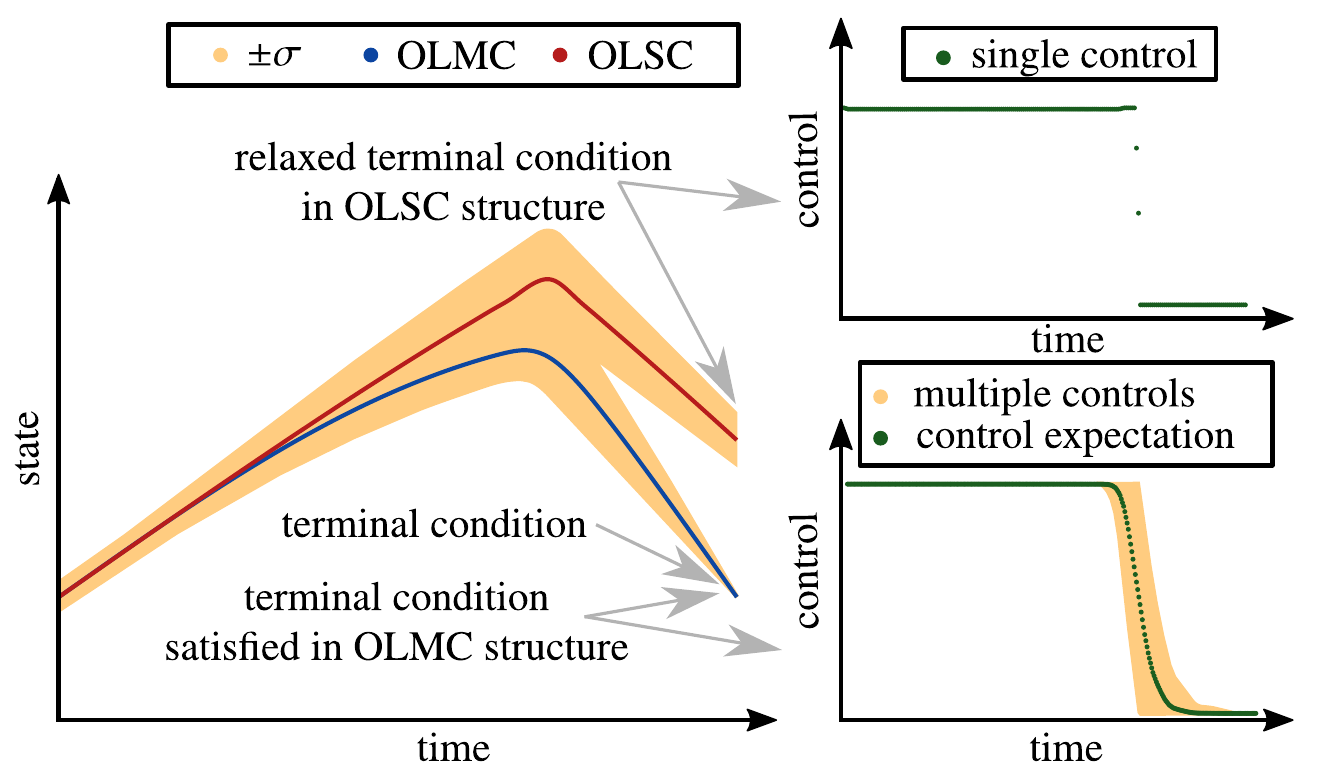}
\caption{Illustration of OLMC, and OLSC structures for an arbitrary problem.}
\label{fig:SCvsMC}
\end{figure}   

This issue has been dealt with in two different ways in the literature: (1) relaxing the prescribed terminal boundary conditions~\cite{azad2020single, azad2021robust}, or (2) minimizing the variance of the terminal state in a multi-objective optimization problem \cite{fisher2011optimal, boutselis2016stochastic}.
The premise for both of these approaches is the assumption that terminal conditions are of Type II equality constraints (i.e.,~the strict satisfaction of such constraints under uncertainty cannot be guaranteed)~\cite{Azad2022, mattson2003handling, rangavajhala2007challenge}.
These remedies enable a solution to the OLSC-UCCD problem, but they do not enforce the terminal boundary conditions.
As a result, the relaxed UCCD problem is an inherently different problem to solve.
This issue is illustrated in Fig.~\ref{fig:SCvsMC}, where the boundary value is relaxed for the OLSC structure to enable a solution. 
In addition, in many real-world applications, relaxing the boundary conditions is not practically viable.
For example, in the UCCD problem of a military aircraft, the vehicle must reach the target or depot despite all uncertainties. 
Therefore, relaxing the terminal condition has the potential to change the meaning and interpretation of the problem, and thus, care must be taken at the time of implementation.
Despite these limitations, OLSC remains a valuable tool in constructing open-loop trajectories for a wide range of reference-tracking applications.  

In terms of implementation, OLSC generally requires the relaxation of some or all of the terminal boundary conditions. 
For a general UCCD problem, a nested coordination strategy may be leveraged to deal with plant uncertainty only in the outer-loop optimization problem.
However, the inner-loop optimal control subproblem will remain coupled through control trajectories\footnote{This is because the single control command is shared among all of the resulting uncertain plants.\vspace{2em} } \textemdash{} resulting in a problem that is generally too large to solve unless a suitable UP method, such as the most-probable-point (MPP), is used to limit the size of the inner-loop problem~\cite{azad2020combined, azad2020single}.  

One remedy for dealing with this issue is to use a DSS approach, where a parameterization of the open-loop control trajectories are the decision variables~\cite{rao2009survey, kelly2017introduction, biegler2010nonlinear, Kelly2017}.
DSS has the advantage of effectively separating control variables from dynamic equations, resulting in sets of dynamic equations that have sparse Jacobian structures and can take advantage of parallel computation.

\subsection{Simple Bounds in UCCD}
\label{subsec:PB}
 
The solution to a UCCD problem might become practically infeasible when simple bounds do not represent the true decision space in the presence of uncertainties \cite{azad2020robust, parkinson1993general}.
To illustrate this further, consider a plant optimization variable $p_{p}$ that is bounded ${p}_{p_{min}} \leq p \leq {p}_{p_{max}}$.
Due to uncertainty-contributing factors, such as imperfect manufacturing processes, the bound on the associated uncertain plant variable $\tilde{p}_{p}$ needs to be modified to reflect the true decision space.
Assuming a Gaussian distribution, a natural way to adjust these bounds is to include a constraint shift index $k_{s}$ such that:
\begin{equation}
    {p}_{p_{min}} + k_{s}\sigma_{p} \leq \tilde{p}_{p} \leq {p}_{p_{max}}-k_{s}\sigma_{p}
\end{equation}

\noindent
where $k_{s}$ is chosen by the designer to reflect the risks associated with these simple bounds. 
As an example, when $k_{s}=3$ and $\tilde{p}_{p}$ has a normal distribution, then the simple bound is satisfied with the probability of $99.865\%$.
This approach employs a crisp uncertainty representation; therefore, it does not always offer a probabilistic interpretation \cite{Azad2022}.

\xsection{Uncertainty Propagation Methods}
\label{sec:section3}
Uncertainty propagation (UP) is a term applied to a family of methods that quantify uncertainties in a system's response and is a critical step in UCCD solution strategies. 
UP methods have been developed for different needs in various research communities, and thus, it is important to understand their utility, merits, and limitations in the context of UCCD.
In a broad classification, UP may be performed in a forward or inverse manner~\cite{liu2018forward, litvinenko2013inverse}. In forward UP, uncertainties from inputs are propagated through the system model, resulting in predictions of system behavior in the presence of uncertainties. 
Inverse UP, on the other hand, uses experimental measurements or observations to estimate discrepancies between the model and the actual system.
Parameter calibration and bias correction are then performed in order to update the model based on such observations. 
While inverse UP is an important step in developing accurate models, it is often the case that actual observations and experimentation are not viable options in early-stage design.
In addition, while a concurrent forward and inverse UP algorithm has the potential to improve the dynamic model and hence, the interpretation of UCCD solutions, this gain comes at the cost of a high computational burden.  
For these reasons, this article only considers methods for forward propagation.

Forward UP methods can be divided into probabilistic and non-probabilistic approaches. 
Probabilistic methods provide an estimate of the likelihood of an event taking place.
This class includes sampling methods such as MCS \cite{fishman2013monte}, local expansion methods such as first-order second-moment (FOSM) \cite{azad2020robust} and perturbation \cite{liu1986probabilistic, stefanou2009stochastic, holmes2012introduction}, functional expansion methods such as gPC \cite{ghanem2003stochastic, xiu2009fast} and Karhunen–Lo\a`eve expansion \cite{gunzburger2011error}, MPP-based methods such as first-order reliability method (FORM) \cite{azad2020single, du2004sequential, youn2005enriched} and second-order reliability method (SORM) \cite{zhang2010second}, and numerical integration methods such as full-factorial numerical integration \cite{lee2009comparative, seo2002efficient}.
Non-probabilistic methods, such as interval analysis or methods based on fuzzy programming, on the other hand, do not provide this information, and their usage is generally limited to conditions where sufficient data is not available \cite{huyse2001random}. 
In addition, the time-evolution of the joint probability distribution function (PDF) in an uncertain system can be estimated directly using direct quadrature method of moments \cite{mcgraw1997description}, direct evolution through Fokker-Planck-Kolmogorov equation \cite{kumar2006multi}, and the Ricatti equation (for linear Gaussian systems) \cite{simon2006optimal}. 
In this section, we limit our discussion to MCS and gPC and discuss their utility and implementation challenges for UCCD problems.

\subsection{Monte Carlo Simulation (MCS)}
\label{subsec:MCS}
UP methods that are based on sampling provide an intuitive way of propagating and quantifying uncertainties in many engineering problems.
These methods generally require a lot of empirical information.
For example, not only the mean value and variance but also the complete probability distribution of uncertain variables must be known for an effective sampling of uncertainties \cite{zio2013literature}.

The sampling methods encompass a wide range of approaches, among which MCS is the most well-known. 
In MCS, random samples of uncertain variables are generated from their joint PDF, and the model is evaluated repeatedly for these samples. 
Therefore, at the root of methods such as MCS is the availability of distributional information for \textit{uncertain basic quantities}.
In this context, uncertain basic quantities are variables described as $\tilde{\bm{x}} \in \mathbb{R}^{n_{x}}$ that have a known PDF whose uncertainty will be propagated into the system. 
As an example, when the uncertainty in an arbitrary UCCD problem stems only from plant optimization variables, then $n_{x} = n_{p}$. 
Using $\tilde{\bm{x}}$ to describe the source of uncertainties will allow us to simplify some of the notations in the remainder of the article.
An unbiased estimator for the mean value of the objective function using MCS can be calculated as:
\begin{equation}
\label{Eqn:MCS_mu}
     {\hat{\mu}_{o}} = \frac{1}{N_{\textrm{mcs}}} \sum_{j=1}^{N_{\textrm{mcs}}}o( t, \bm{u}_{j}, \bm{\xi}_{j}, \bm{p}_{j}, \pd_{j})
\end{equation}

\noindent
where $N_{\textrm{mcs}}$ is the number of samples. 
The variance can be estimated as:
\begin{equation}
\label{Eqn:MCS_sigma2}
     {\hat{\sigma}_{o}^{2}} = \frac{1}{N_{\textrm{mcs}}-1} \sum_{j=1}^{N_{\textrm{mcs}}}\left (o( t, \bm{u}_{j}, \bm{\xi}_{j}, \bm{p}_{j}, \pd_{j}) - {\hat{\mu}_{o}} \right )^2
\end{equation}

\noindent
Equations (\ref{Eqn:MCS_mu})--(\ref{Eqn:MCS_sigma2}) can now be used in Eq.~(\ref{Eqn:UCCD}) to produce any of the specialized formulations based on stochastic and probabilistic robust formulations \cite{Azad2022}. 

MCS is flexible, easy to implement, and capable of offering high solution accuracy (up to a certain limit at high cost).
Therefore, it is often used to benchmark newly-developed methods.
However, a prohibitively large number of samples are required to estimate rare events accurately (i.e.,~events with $\mathbb{P}\ll 1$).
The convergence rate of the expected value in MCS is $\mathcal{O}(1/\sqrt{N_{\textrm{mcs}}})$ \cite{xiu2009fast, fishman2013monte}. 
Interestingly, this convergence rate does not depend on the dimension $n_{x}$, which is an advantage when compared to some of the other UP methods.

\subsection{Generalized Polynomial Chaos (gPC)}
\label{subsec:gPC}

\begin{figure*}[t]
\centering
\includegraphics[width=\textwidth]{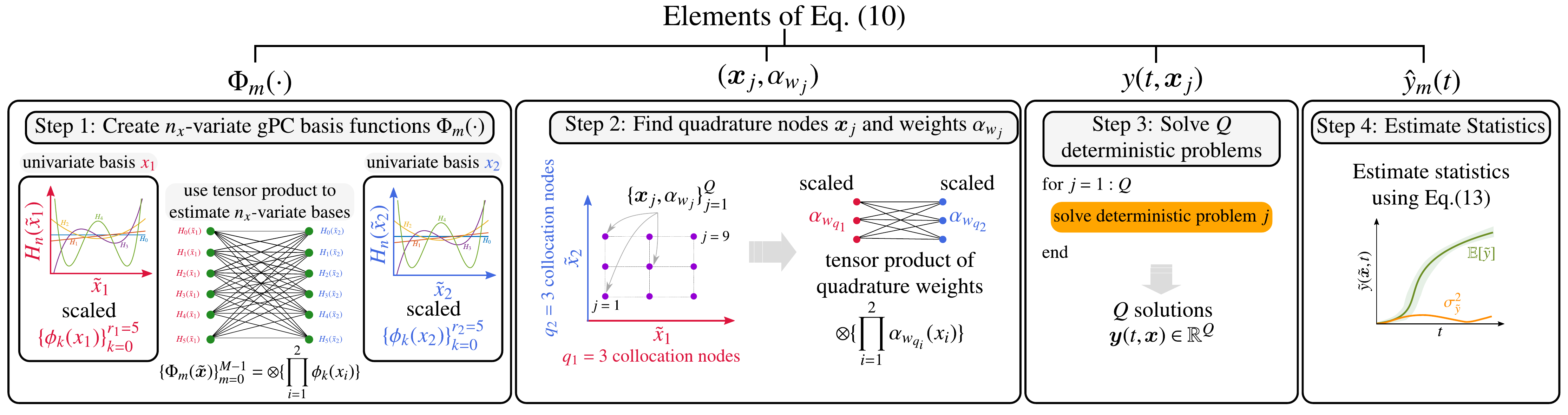}
\caption{Illustration of steps involved in uaing gPC for uncertainty propagation in a two-dimensional uncertain problem: Step 1: Create the $n_x$-variate gPC basis functions $\Phi_{m}(\cdot)$ using the tensor product of univariate basis functions selected from Table~\ref{Tab:Polyrep}, Step 2: Find quadrature nodes $\bm{x}_{j}$ and weights $\alpha_{w_{j}}$ by selecting a certain number of collocation nodes in each dimension, Step 3: solve the deterministic problem at all of the quadrature nodes to find $y(t,\bm{x}_{j})$ where ${y}$ is any desired uncertain output such as states or controls, and Step 4: estimate statistics of the output using Eq.~(\ref{Eq:gPC9}).}
\label{fig:gPC}
\end{figure*} 

Generalized polynomial chaos represents stochastic variables and processes using orthogonal polynomials. 
Consider a random vector $\tilde{\bm{x}}$ with mutually independent components and distribution $F_{\tilde{\bm{x}}}(\bm{x}) = \mathbb{P}(\tilde{x}_{1} \leq x_{1}, \dots, \tilde{x}_{n_{x}} \leq x_{n_{x}})$.
If the random elements in $\tilde{\bm{x}}$ are not independent, then a Karhunen–Lo\a`eve decomposition \cite{loeve1978probability} or a Rosenblatt transformation \cite{rosenblatt1952remarks} should first be applied.
In generalized polynomial chaos, we use the tensor product of appropriately-selected univariate orthogonal basis functions of degree up to $r_{i}$, denoted as $\{\phi_{k}(\tilde{x}_{i})\}_{k=0}^{r_{i}}$ in order to construct the $n_{x}$-variate gPC basis functions of degree $M$ as $\{{\Phi}_{m}(\tilde{\bm{x}})\}_{m=0}^{M-1}$ (Step 1 of Fig.~\ref{fig:gPC}).
If we keep the highest polynomial order for up to $PC$ in each direction, then $\vert \bm{k} \vert = \max_{1\leq j \leq n_{x}} k_{j} $ and the resulting polynomials have the dimension of $M =\prod_{i=1}^{n_{x}}(PC+1)$. 
Alternatively, a subset of basis elements with a total degree of up to $PC$ can be selected.
In this case, the $n_{x}$-variate gPC is defined using a multi-index $\bm{k} = (k_{1},\dots,k_{n_{x}}) \in \mathbb{N}_{0}^{n_{x}}$, where $\mathbb{N}_{0}^{n_{x}}$ is the set of $n_{x}$-dimensional natural numbers with zero, and $\bm{k}$ is a multi-index with $\vert \bm{k} \vert = k_{1}+ \dots + k_{n_{x}}$.
The resulting polynomials have the dimension of $M$:
\begin{equation}
    M = \binom{PC+n_{x}}{PC}
\end{equation}

\noindent
The $n_{x}$-variate gPC basis functions then allow us to estimate any general second-order variable or process $\tilde{y}(t,\tilde{\bm{x}})$ as:
\begin{equation}
\label{Eq:gPC6}
    \begin{aligned}
    \tilde{y}(t,\tilde{\bm{x}}) \approx y_{PC}(t,\tilde{\bm{x}}) = \sum_{m=0}^{M-1} \hat{y}_{m}(t){\Phi}_{m}(\tilde{\bm{x}})
    \end{aligned}
\end{equation}

\noindent
where $y_{p_{c}}(\cdot)$ is the $PC$th-degree gPC approximation of $\tilde{y}(\cdot)$, and $\hat{y}_{i}(t)$ are unknown coefficients.
In the context of UCCD, any uncertain problem element (such as $\tilde{\bm{u}}$, $\tilde{\bm{\xi}}$, $\tilde{\bm{p}}$, etc.) can be approximated using this approach.

From here, we can use either a Galerkin or a collocation formulation of gPC to calculate the unknown coefficients $\hat{y}_{m}(t)$ in Eq.~(\ref{Eq:gPC6}).
The Galerkin formulation, which is an intrusive UP method, uses a Galerkin projection on the basis functions to approximate uncertain quantities.
The development and application of the Galerkin type of gPC can be found in Refs.~\cite{xiu2010numerical, wang2019robust, boutselis2016stochastic, fisher2011optimal}.
In the collocation formulation of gPC, the $m$th unknown coefficient, $\hat{y}_{m}(t)$ can be obtained using:
\begin{equation}
\label{Eq:gPC7}
    \begin{aligned}
    & \hat{y}_{m}(t) = \mathbb{E}\left[y(t,\bm{x}) \Phi_{\bm{j}}(\tilde{\bm{x}})\right] = \int_{\Gamma} y(t,\bm{x})\Phi_{m}(\bm{x})dF_{\tilde{x}}(\bm{x}) \\
    \end{aligned}
\end{equation}

\noindent
where $\Gamma$ is the finite domain of the distribution function, constructed from the product of independent finite domains associated with each uncertain dimension, i.e.,~ $\Gamma = \prod_{i=1}^{n_{x}}\Gamma_{i}$.
The evaluation of this integral requires a quadrature rule, and thus, a set of collocation nodes, along with their associated quadrature weights, must be selected according to the corresponding polynomial representation in Table~\ref{Tab:Polyrep}.
We define $q_{i}$ user-defined collocation nodes in each $i$ dimension of uncertain quantities.
The full tensor product of these user-defined nodes can then be used to generate an $n_{x}$-dimensional grid of $Q$ collection nodes and quadrature weights $\alpha_{w}$, such that $\{\bm{x}_{j},\alpha_{w_{j}} \}_{j=1}^{Q}$, where $Q = \otimes \{q_{1}, \dots, q_{n_{x}}\}$ (Step 2 of Fig.~\ref{fig:gPC}).
Using the quadrature rule, Eq.~(\ref{Eq:gPC7}) can be written as:
\begin{equation}
\label{Eq:gPC8}
    \begin{aligned}
    & \hat{y}_{m}(t) = \sum_{j=1}^{Q} y(t,\bm{x}_{j})\Phi_{m}(\bm{x}_{j})\alpha_{{w}_{j}} ~~ m = 0,\ldots, M \\
    \end{aligned}
\end{equation}

\noindent
where $\bm{x}_{j}$ is the $j$th collocation node and $\alpha_{w_{j}}$ is the associated quadrature weight.
Note that now, $y(t,\bm{x}_{j})$ is the deterministic solution to the $j$th sample of random vector (Step 3 of Fig.~\ref{fig:gPC}). 
When an accurate gPC for an arbitrary random function $\tilde{y}(t,\tilde{\bm{x}})$ is constructed, the function can be analytically represented as a function of $\tilde{\bm{x}}$. 
Therefore, all statistical information of $\tilde{y}(\cdot)$ can be estimated with little computational effort \cite{xiu2010numerical}. 
Statistics of $\tilde{y}(t, \tilde{x})$ are then calculated as (Step 4 of Fig.~\ref{fig:gPC}):
\begin{equation}
\label{Eq:gPC9}
    \begin{aligned}
    & \mathbb{E}\left[\tilde{y}(t, \tilde{x})\right] \approx \mathbb{E}[\tilde{y}_{PC}(t, \tilde{x})] = \hat{y}_{1}(t)  \\
    &\sigma_{\tilde{y}(t, \tilde{x})}^{2} \approx \sigma_{\tilde{y}_{PC}(t, \tilde{x})} = \sum_{m=2}^{M} \hat{y}_{m}^{2}(t) 
    \end{aligned}
\end{equation}

\begin{table}[t]
    \caption{Correspondence between the type of gPC and their underlying random variables for select distributions \cite{xiu2010numerical}.}
    \label{Tab:Polyrep}
    \renewcommand{\arraystretch}{1.1}
    \centering
    \begin{tabular}{c c c c }
    \hline \hline
    & \textrm{\textbf{Distribution}}  & \textrm{\textbf{gPC polynomial}} & \textrm{\textbf{Support}}  \\
    \hline
    \multirow{4}{*}{\rotatebox[origin=c]{90}{Continuous}} & \multicolumn{1}{|c}{\textrm{Gaussian}} & Hermite &  $(-\infty, \infty)$ \\
     & \multicolumn{1}{|c}{\textrm{Gamma}} & Laguerre & $[0,\infty)$ \\
     & \multicolumn{1}{|c}{\textrm{Beta}} & Jacobi   &  {$[a,b]$}\\
     & \multicolumn{1}{|c}{\textrm{Uniform}} & Legendre   &  $[a,b]$  \\
    \hline
    \multirow{4}{*}{\rotatebox[origin=c]{90}{Discrete}}& \multicolumn{1}{|c}{\textrm{Poisson}} & Charlier &  $\{0,1,\dots\}$ \\
    & \multicolumn{1}{|c}{\textrm{Binomial}} & Krawtchouk & $\{0,1,\dots,N_{d} \}$ \\
    & \multicolumn{1}{|c}{\textrm{Negative binomial}} & Meixner   & $\{0,1,\dots\}$  \\
     & \multicolumn{1}{|c}{\textrm{Hypergeometric}} & Hah   &   $\{0,1,\dots,N_{d} \}$ \\
    \hline \hline
    \end{tabular}
\end{table}

\xsection{Simple SASA UCCD Formulations}
\label{sec:section4}

\begin{figure}[t]
\centering
\includegraphics[scale=0.55]{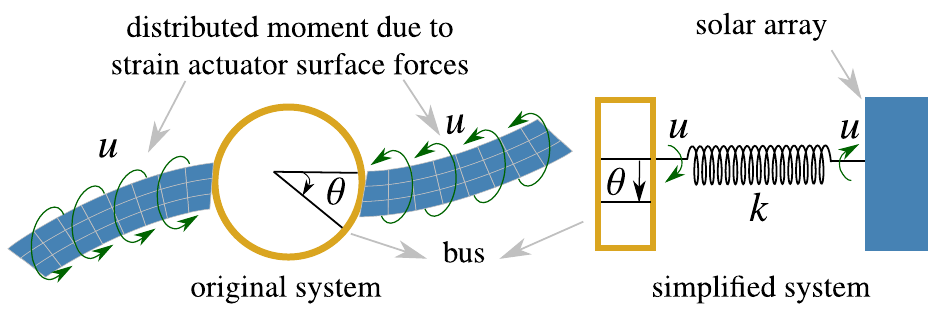}
\caption{Illustration of the original and simplified strain-actuated solar array systems. }
\label{fig:SimpleSASA}
\end{figure}

In this section, we construct the SE-UCCD and WCR-UCCD formulations to investigate the integrated UCCD solution of the simple SASA case study as a means to explore UCCD more broadly. 
Modified from Refs.~\cite{herber2017unified, chilan2017co}, this problem describes a simplified version of a strain-actuated solar array (SASA) system for spacecraft pointing control and jitter reduction. 
As shown in Fig.~\ref{fig:SimpleSASA}, this system uses distributed actuators to strain the solar arrays, which results in reactive forces that can be used to control the spacecraft body.
The objective is to maximize the displacement of the spacecraft's body at the final time.
The deterministic simple SASA problem is formulated as:

\begin{subequations}
 \label{Eqn:SASA_det}
 \begin{align}
\underset{u,\bm{\xi}, k}{\textrm{minimize:}}
\quad &  -\xi_{1}(t_f)   \label{SASA_det_obj} \\
\textrm{subject to:} \quad
& u - u_{max} \leq 0 \label{SASA_det_ineq1} \\
& u_{min} - u \leq 0 \label{SASA_det_ineq2} \\
&\begin{bmatrix}
& \dot{\xi_{1}} \\ 
& \dot{\xi_{2}} 
\end{bmatrix}
=
\begin{bmatrix}
0 & 1 \\
-\frac{k}{J} & 0
\end{bmatrix}
\begin{bmatrix}
& \xi_{1} \\ 
& \xi_{2} 
\end{bmatrix}
+ 
\begin{bmatrix}
0\\
\frac{1}{J}
\end{bmatrix}
u \label{SASA_det_dyn} \\
&k \xi_{1}(t_{f}) -u_{max} \leq 0 \label{SASA_det_stationary}\\
&\bm{\xi}(t_0) = \begin{bmatrix}
0\\
0
\end{bmatrix},~ 
\xi_{2}(t_f) = 0 \label{SASA_det_bc} \\
\textrm{where:} \quad  & u(t) = u, ~ {\bm{\xi}}(t) = {\bm{\xi}}  \label{SASA_det_where}
\end{align}
\end{subequations}

\noindent
In this problem, $u$ is the open-loop control moment applied to the solar array, ${\xi}_{1}$ and $\xi_{2}$ are state variables (associated with the relative displacement and velocity), and $k$ is the plant optimization variable associated with the stiffness of the solar array.
The inertia ratio between the solar array and the bus is described by $J$, and considered problem data.
Equation (\ref{SASA_det_stationary}) represents the final condition that enables the system to return to the stationary condition.
The number of optimization variables is $n_{u} = 1$, $n_{s}=2$, and $n_{p}=1$.
Note that $u$ is bounded by $u_{min}$ and $u_{max}$, and the initial and terminal boundary conditions are imposed through Eq.~(\ref{SASA_det_bc}).
For the simplicity of notation when introducing the associated UCCD formulations, we define the feasible set of Type~I equality constraints as $\mathcal{E}$:
\begin{align}
    \label{eqn:equalityFeasibility}
    \mathcal{E} = \{ (t,\bm{u}, \bm{\xi}, k, J, {\bm{\xi}(t_{0})}) \mid \dot{\bm{\xi}}(t) = \bm{f}(\cdot) \}
\end{align}

\noindent
where $ \bm{f}(\cdot)$ is the functional relationship defined in Eq.~(\ref{SASA_det_dyn}).

When uncertainties are present, depending on the availability of distributional information, the simple SASA UCCD problem may be formulated like any of the specialized forms discussed in detail in Ref.~\cite{Azad2022}.     
In this article, we assume that uncertainties originate from plant optimization variable $\tilde{k}$, uncertain problem data $\tilde{J}$, and uncertain initial boundary condition for the second state variable $\tilde{\xi}_{2,t_{0}}$. 
Generally, time-dependent disturbances are also present in the dynamic model and may require specific treatment before they can be efficiently used in gPC. 
For this reason, this study only focuses on time-independent uncertainties.
We also assume that all of these uncertain quantities have a Gaussian distribution with known standard deviations of $\sigma_{k}$, $\sigma_{J}$, and $\sigma_{\xi_{2,t_{0}}}$, respectively.
Note, however, that other distributions may also be used for gPC according to Table~\ref{Tab:Polyrep}.
In addition, for the crisp representation of uncertainties, we assume that uncertainties belong to bounded, deterministic sets constructed within the range of $\pm k_{s} \sigma$ of the mean values of uncertain quantities, where $k_{s} =3$.
The stochastic and crisp representations of these uncertainties are described in Table~\ref{tab:Uncertainties}.
 
Since uncertainties are represented in a probabilistic and crisp manner, three distinct formulations become viable: SE-UCCD, PR-UCCD, and WCR-UCCD \textemdash{} all of which can be derived from the universal UCCD formulation in Eq.~(\ref{Eqn:UCCD}) \cite{Azad2022}.
In this article, we only focus on SE-UCCD and WCR-UCCD implementations.
For the simple SASA problem, it is important to bring the vehicle to rest at the final time.
This condition may be necessary for accurate positioning, safety, and functionality of the system.
Therefore, the simple SASA problem readily lends itself to the OLMC and MSC structures.
Thus, in the first step, this article implements and solves the OLMC-SE-UCCD problem using MCS and gPC.
Next, using polytopic uncertainties along with some results from linear programming theory, we implement an OLMC-WCR-UCCD formulation of the simple SASA problem.
An MSC-WCR-UCCD is also implemented to complement the results and provide insights into the value of feedback information, as well as the role of control structure in managing risk versus performance.
Finally, such investigations are supplemented by performing close-loop feedback control analyses of systems' designs to gain more insights into systems' behavior.

\begin{table}[t]
    \caption{Summary of uncertain quantities.}
    \label{tab:Uncertainties}
    \renewcommand{\arraystretch}{1.2}
    \centering
    \begin{tabular}{l c c c c }
    \hline \hline
    $\tilde{\bm{q}}$ & $\mu_{q}$ & $\sigma_{q}$  & \text{Stochastic} & {Crisp}  \\
    \hline
    $\tilde{k}$ & $\mu_{k}$  & $0.20$ & $\mathcal{N}(\mu_{k},\sigma_{k})$ & [$\mu_{k} \pm 0.6 $] \\
    $\tilde{J}$ & $\mu_{J} = 1$  & $0.15$ & $\mathcal{N}(\mu_{J},\sigma_{J})$ & $[0.55~~ 1.45]$ \\
    $\tilde{\xi_{2}}(t_0)$ & $\mu_{\xi_{2,t_{0}}} = 0$  & $0.03$ & $\mathcal{N}(\mu_{\xi_{2,t_{0}}},\sigma_{\xi_{2,t_{0}}})$ &  $[-0.09~~0.09]$\\
    \hline \hline
    \end{tabular}
\end{table}

\subsection{Stochastic in Expectation (SE-UCCD)}
\label{PUCCD Example1}

When distributional information of uncertainties is available, the deterministic CCD problem may be formulated as SE-UCCD.
In this formulation, the objective function and inequality constraints are modeled in terms of their expected values.
This points to the \textit{risk-neutral nature} of this formulation because no measures are taken to reduce the risks associated with constraint violation.
The OLMC structure solves the problem subject to all boundary conditions by eliciting a distinct control response for every realization of uncertainties.  
The simultaneous SE-UCCD problem is formulated as:
\begin{subequations}
 \label{Eqn:SASA_PUCCD}
 \begin{align}
 \underset{u,\tilde{\bm{\xi}}, \mu_{k}}{\textrm{minimize:}}
 \quad & -\mathbb{E}[\tilde{\xi}_{1}(t_f)]  \label{SASA_PUCCD_obj} \\
 \textrm{subject to:} \quad
& \textrm{Eqs.~ (\ref{SASA_det_ineq1})-(\ref{SASA_det_ineq2}}) \label{SASA_PUCCD_ineq}\\
& {(t,u,\tilde{\bm{\xi}},\tilde{k},\tilde{J},{\bm{\xi}(t_{0})}) \in \mathcal{E}} \label{SASA_PUCCD_dyn}  \\
& k_{s}\sigma_{k} - {\mu}_{k} \leq 0 \label{SASA_PUCCD_plant}\\
&{ \mu_{k} \tilde{\xi}_{1}(t_{f}) -u_{max} \leq 0 } ~~ (\textrm{if OLMC or MSC}) \label{SASA_PUCCD_stationary}\\
&\tilde{\bm{\xi}}(t_0) = \begin{bmatrix}
0\\
\mathcal{N}(\mu_{\xi_{2,t_{0}}}, \sigma_{\xi_{2,t_{0}}})
\end{bmatrix}\label{SASA_PUCCD_bc1}\\
&\tilde{\xi}_{2}(t_f) = 0 ~~ (\textrm{if OLMC})
\label{SASA_PUCCD_bc2}\\
\textrm{where:} \quad  & \tilde{k}= \mathcal{N}(\mu_{k},\sigma_{k}), ~~ \tilde{J} = \mathcal{N}(\mu_{J}, \sigma_{J}) \label{SASA_PUCCD_where1}
\end{align}
\end{subequations}

\noindent
In this equation, $\mathcal{N}(\cdot)$ describes a normal distribution, $\mu_{k}$, $\mu_{J}$, and $\mu_{\xi_{2,t_{0}}}$ are the mean values of the solar array stiffness, inertia ratio, and the initial condition for $\xi_{2}$, respectively.
Equation (\ref{SASA_PUCCD_dyn}) describes uncertain system dynamics, which are of Type I equality constraints \cite{Azad2022}.
Therefore, they must be satisfied ``almost surely'' or ``a.s.'' (with the probability of one).
That is why these equations are described in terms of the infinite-dimensional uncertain quantities such as $\tilde{k}$ and $\tilde{J}$. 
Equation~(\ref{SASA_PUCCD_plant}) is included to ensure the feasibility of any boundary solution according to the discussion in Sec.~\ref{subsec:PB}. 
In this equation, $k_{s}$ is the constraint shift index.
Equation (\ref{SASA_PUCCD_stationary}) enables the system to go to stationary conditions, which is only enforced for OLMC and MSC structures.
Similarly, Eq.~(\ref{SASA_PUCCD_bc2}) is only enforced for OLMC structures.

\subsection{Multi-Stage Control}
\label{MPC}
Using polytopic uncertainties (which are described in detail in Sec.~\ref{WCUCCD Example1}), the MSC-UCCD problem can be implemented by replacing the inner-loop optimal control subproblem (Eq.~(\ref{Eqn:SASA_WCUCCD_MC})) with a sequence of receding-horizon problems.
Due to the system's linearity in the nested coordination strategy, and for practical purposes, all uncertain plants were used to create one large dynamic system. 
In addition, for a robust implementation (i.e.,~avoiding infeasible outcomes from the sequence of inner-loop sub-problems), the terminal condition was removed from the set of constraints and posed as a quadratic penalty term in the objective function.

\subsection{Worst-Case Robust (WCR-UCCD)}
\label{WCUCCD Example1}
When the probabilistic information of uncertain quantities is not available, a worst-case robust formulation provides a conservative solution that is optimal for the worst-case realization of uncertainties within the uncertainty set.
Therefore, a major assumption in implementing such a formulation is the knowledge of the geometry and size of the uncertainty set.
We assume that uncertainties belong to the following uncertainty set:
\begin{align}
    \label{uncertaintyseta}
    \mathcal{S} \coloneqq  & \left \{ \bm{q}~\mid \bm{z} \left( \hat{\bm{q}} - \bm{q} \right )  \leq \bm{\eta}_{q} \right \}
\end{align}

\noindent
where $\mathcal{S}$ is the uncertainty set, and $\bm{q}$ is a column vector composed of the concatenation of all uncertain quantities, and $\hat{\bm{q}}$ are nominal quantities that are used to construct the uncertainty sets.
For the simple SASA problem, ${\bm{q}}^{T} \coloneqq [k, J,\xi_{2,t_{0}} ]$ and $\hat{\bm{q}}^{T} \coloneqq [\mu_k, \mu_J,\mu_{\xi_{2,t_{0}}} ]$.
Furthermore, $\bm{z}$ is a specified function chosen to represent the geometry of the uncertainty set, often through an applied norm such as $\ell_{1}$, $\ell_{2}$, $\ell_{p}$, $D$, CVaR, etc. \cite{rahal2021norm}, and $\bm{\eta}_{q}$ determines the size of the uncertainty set.
Since all of the assumed uncertainties in this problem have a normal distribution, a natural choice in constructing the uncertainty set is to assume that $\bm{\eta}_{q} = \pm k_{s}\sigma_{q}$, where $k_{s} = 3$, with $\bm{z}$ being an $\ell_{\infty}$ norm.
The epigraph representation of the worst-case robust formulation for the outer loop of the simple SASA problem is formally described as \cite{Azad2022}:
\begin{subequations}
 \label{Eqn:SASA_WCUCCDo}
 \begin{align}
 \underset{u,\mu_{k},v}{\textrm{minimize:}}
 \quad & -v  \label{SASA_WCUCCDo_obj} \\
 \textrm{subject to:} \quad & v - \Phi \left (t,{u},k(\mu_{k}) \right) \leq 0 \label{SASA_WCUCCDo_vobj}   \\
 & {(t,u,\bm{\xi},\mu_{k},\mu_{J},\mu_{\xi_{2,t_{0}}}) \in \mathcal{E}} \label{SASA_WCUCCDo_dyn} \\
 & \textrm{Eqs.~(\ref{SASA_det_ineq1})--(\ref{SASA_det_ineq2})} \label{SASA_WCUCCDo_ineq}
\end{align}
\end{subequations}

\noindent
where $u(t)$ and $\mu_{k}$ are inputs to the inner-loop optimization problem, and $\Phi(\cdot)$ is the original objective function replaced by the new decision variable $v$.
The inner-loop WCR problem is formulated as:
\begin{subequations}
\label{Eqn:SASA_WCUCCDi}
\begin{align}
\underset{{\bm{\xi}}, k, J, \xi_{2,t_{0}}}{\textrm{minimize:}}
\quad & o_{in} = {\xi}_{1}(t_f)  \label{SASA_WCUCCDi_obj} \\
\textrm{subject to:} \quad   & 
(t,u,\bm{\xi}, k,J, {\bm{\xi}(t_{0})}) \in \mathcal{E} \label{SASA_WCUCCDidyn}  \\
& \textrm{Eq.~(\ref{SASA_det_stationary})} ~~ (\text{if OLMC or MSC})
\label{SASA_WCUCCDi_stationary}\\
&{\bm{\xi}}(t_0) = \begin{bmatrix}
0\\
{\xi}_{2,t_{0}}
\end{bmatrix}
\label{SASA_WCUCCDi_bc}\\
&\xi_{2}(t_{f}) = 0 ~~ (\text{if OLMC}) \label{TC} \\
\textrm{where:} \quad  & [k(\mu_{k}), J, \xi_{2,t_{0}}] \in \mathcal{S}
\label{SASA_WCUCCDi_where2}
\end{align}
\end{subequations}

\noindent
The above formulation described the OLSC-WCR-UCCD formulation.
The implementation of the OLMC-WCR-UCCD can be simplified by recognizing that (i) uncertainties are represented as a closed, bounded set and form a convex polytope, and (ii) the OLMC-WCR-UCCD problem is a linear program when implemented within a nested coordination strategy. 
A polytope is defined as a bounded, closed, and convex polyhedron described by a finite number of affine inequalities.
For a linear program, the feasible region of the optimization problem is a convex polytope (i.e.,~the convex hull of the vertices of the polytope), and the optimal solution is achieved at a vertex. 
Therefore, the OLMC-WCR-UCCD only requires the solution of the control co-design problem at the vertices of the polytope. 
Describing the set of polytope vertices by $\mathcal{P}_{d}$, and taking advantage of a nested coordination strategy, the OLMC-WCR-UCCD formulation for the $i$th vertex may be defined as 
\begin{subequations}
 \label{Eqn:SASA_WCUCCD_MC_O}
 \begin{align}
 \underset{\mu_k}{\textrm{minimize:}}
 \quad & -{\xi}_{1}(t_f)  \label{SASA_WCUCCD_MC_obj_o} \\
 \textrm{subject to:} \quad & 
  \textrm{Eq.~(\ref{SASA_PUCCD_plant})} 
\end{align}
\end{subequations}
\noindent 
where the inner-loop optimal control sub-problem is defined as
\begin{subequations}
 \label{Eqn:SASA_WCUCCD_MC}
 \begin{align}
 \underset{\bm{u},{\bm{\xi}}}{\textrm{minimize:}}
 \quad & -{\xi}_{1}(t_f)  \label{SASA_WCUCCD_MC_obj} \\
 \textrm{subject to:} \quad & \textrm{Eqs.~(\ref{SASA_WCUCCDidyn})--(\ref{TC}) \label{SASA_WCUCCD_MC_dyn}}\\
 \textrm{where:} \quad  & [k(\mu_{k}), J, \xi_{2,t_{0}}] = \upsilon_{i} \in \mathcal{P}_{d}
\label{SASA_WCUCCD_MC_whereo}
\end{align}
\end{subequations}
where $ \upsilon_{i}$ is the $i$th element of $\mathcal{P}_{d}$.
For practical purposes, all analysis-type equality constraints are moved into the inner-loop problem.
Once all of the solutions at the vertices of the polytope are obtained, the worst-case robust UCCD solution is the one corresponding to the poorest performance of the system.

\xsection{Results and Discussion}
\label{sec:section5}

The simple SASA UCCD was implemented and solved for the OLMC-SE-UCCD, OLMC-WCR-UCCD, and MSC-WCR-UCCD formulations.
Since a nested UCCD coordination strategy results in linear dynamics and, thus, improved computational cost, this coordination strategy has been consistently used.
All of the inner-loop, optimal control subproblems from the nested coordination strategy are solved through a direct transcription (DT) approach.
For all of the DT implementations, the \textsc{Matlab}-based \textit{DTQP} toolbox was used to efficiently construct and solve the UCCD problem \cite{herber2017advances, DTQP}.
\textit{DTQP} toolbox offers a simple problem definition approach, options for differentiation, and the possibility to take advantage of the linearized dynamics to obtain an efficient solution.
The SE-UCCD formulation was solved using MCS and gPC. 
The code, along with all of the results and implementations, are made available in Ref.~\cite{code}.

\begin{table}[t]
    \caption{Settings for UCCD implementations.}
    \label{Tab:CS1_settings}
    \renewcommand{\arraystretch}{1.1}
    \centering
    \begin{tabular}{l l l l}
    \hline  \hline
    \textrm{\textbf{Category}} & \textrm{\textbf{Option}} & \textrm{\textbf{Value}} & \\
    \hline
    \multirow{5}{*}{\textrm{General}} & \textrm{defects} & \textrm{trapezoidal (TR)} \\
    & \textrm{mesh} & \textrm{equidistant} & \\
    & \textrm{quadrature} & \textrm{composite TR} & \\
    & \textrm{outer-loop solver} & \textrm{fmincon} & \\
    & \textrm{solver tolerance} & $10^{-6}$ & \\
    \hline
    \multirow{7}{*}{\textrm{SE-UCCD}}& \textrm{inner-loop solver} & \textrm{quadprog} & \\
    & \textrm{derivatives} & \textrm{symbolic} & \\
    & $n_{t}$ & $100$ & \\
    & $N_{mcs}$ & $10,000$ & \\
    & $Q$ & $10^{3}$ & \\
    & $r_{i}$ & $8$ & \\
    & $M$ & $9^{3}$ & \\
    \hline
    \multirow{3}{*}{\textrm{WCR-UCCD}} & \textrm{inner-loop solver} & \textrm{fmincon} \\
    & \textrm{derivatives} & \textrm{forward} & \\
    & $n_{t}$  & $100$ & \\
    \hline \hline
    \end{tabular}
\end{table}

The computer architecture used for these case studies is a desktop workstation with an AMD 3900X 12-core processor at 3.79 \unit{GHz}, 32 \unit{GB} of RAM, 64-bit Windows 10 Enterprise LTSC version 1809, and \textsc{Matlab} R2021b. 
The settings associated with the \textit{DTQP} toolbox, optimization solvers, and specific method-dependent parameters for MCS and gPC are reported in Table~\ref{Tab:CS1_settings}. 
The results from these UCCD implementations for the simple SASA problem are presented in Table~\ref{tab:CS1_results}.

\begin{figure*}[ht]
\captionsetup[subfigure]{justification=centering}
\centering
\begin{subfigure}{0.25\textwidth}
\centering
\includegraphics[scale=0.34]{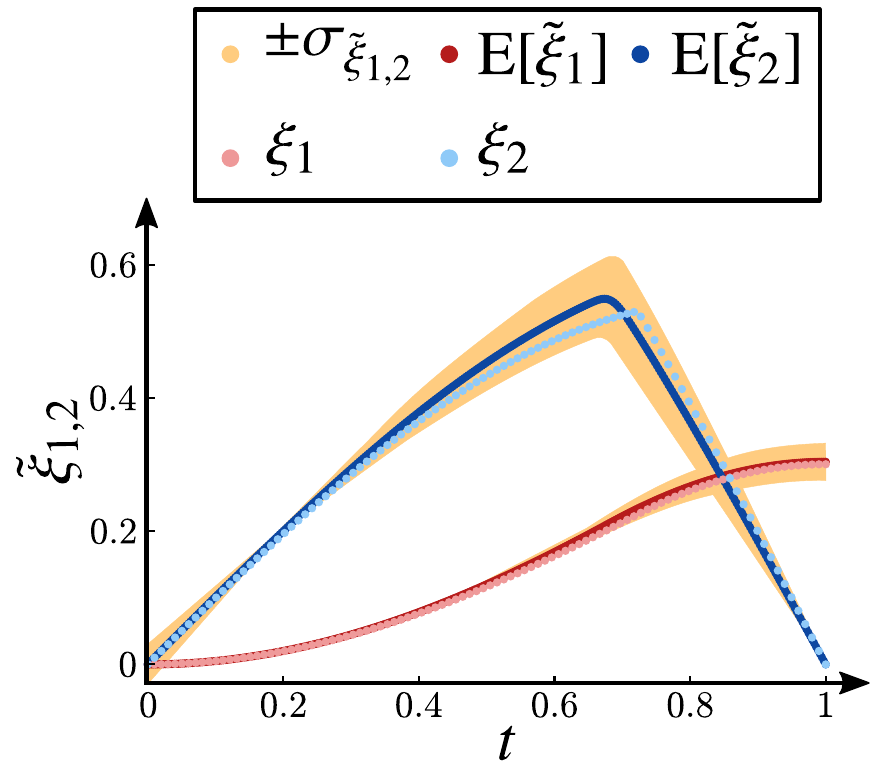}
\caption{Optimal states using MCS. }
\label{fig:STC_MC_MCS_st}
\end{subfigure}%
\begin{subfigure}{0.25\textwidth}
\centering
\includegraphics[scale=0.34]{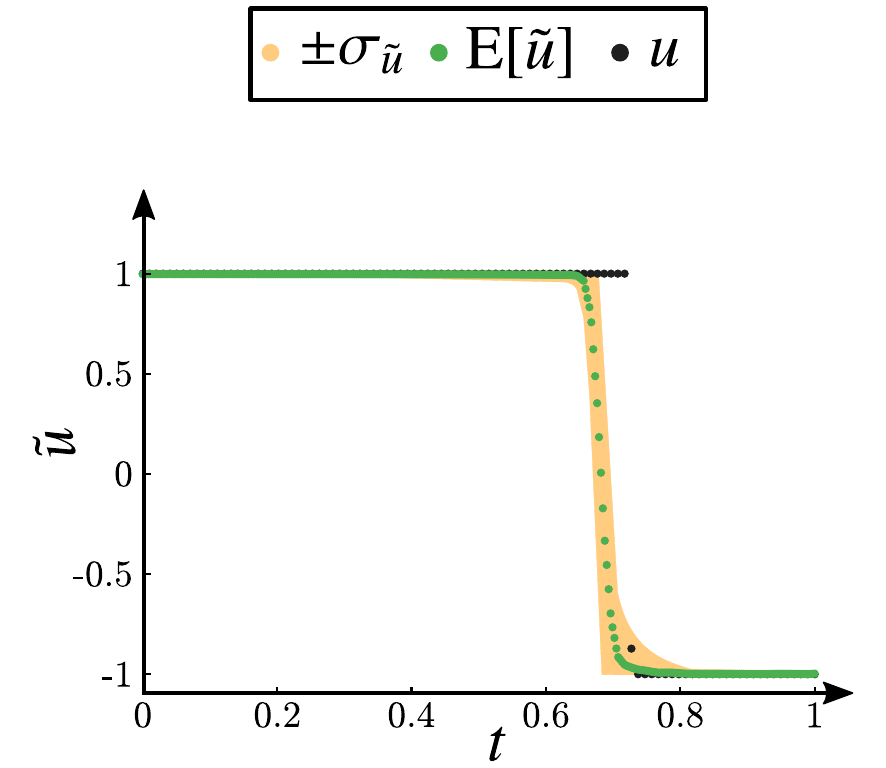}
\caption{Optimal control using MCS.}
\label{fig:STC_MC_MCS_ctrl}
\end{subfigure}%
\begin{subfigure}{0.25\textwidth}
\centering
\includegraphics[scale=0.34]{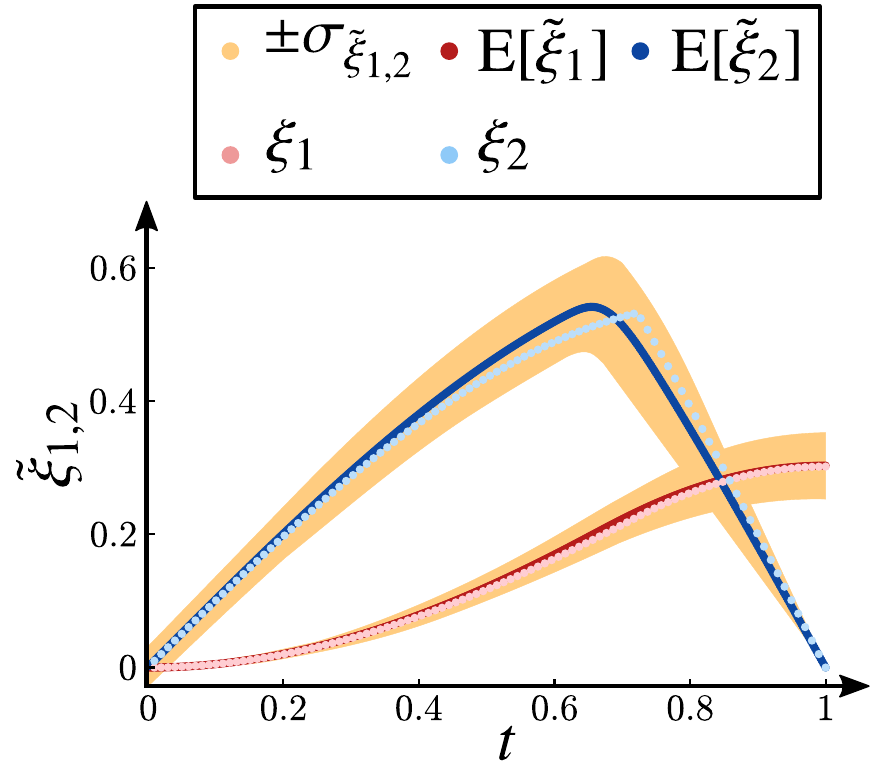}
\caption{Optimal states using gPC.}
\label{fig:STC_MC_gPC_st}
\end{subfigure}%
\begin{subfigure}{0.25\textwidth}
\centering
\includegraphics[scale=0.34]{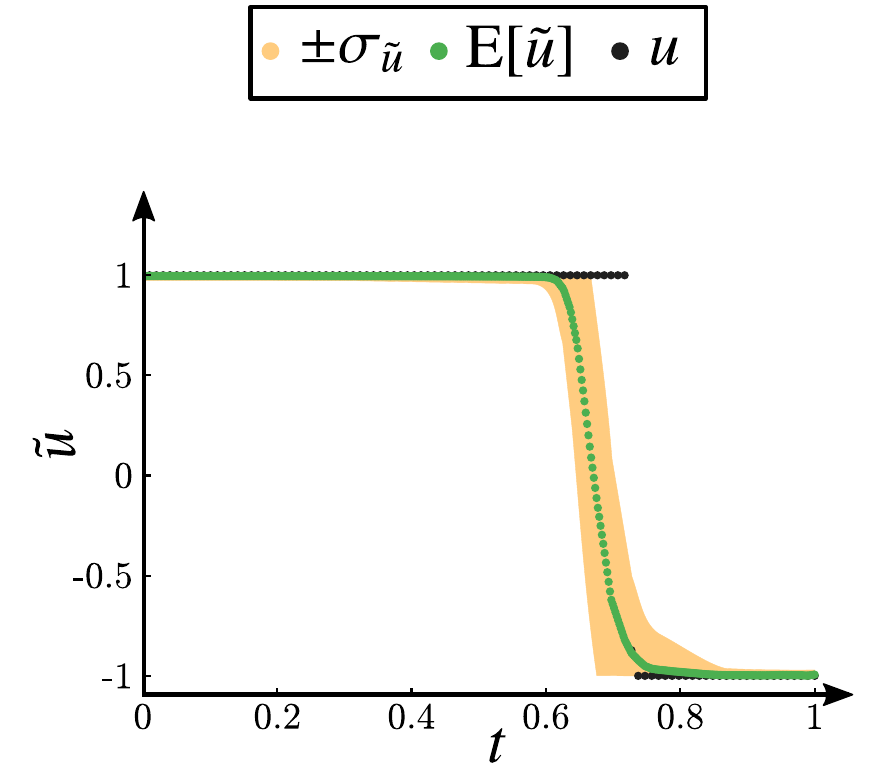}
\caption{Optimal control using gPC. }
\label{fig:STC_MC_gPC_ctrl}
\end{subfigure}
\captionsetup[figure]{justification=centering}
\caption{Open-loop multiple-control (OLMC), stochastic in expectation UCCD (SE-UCCD) solution using MCS and gPC in comparison to the deterministic CCD solution.}
\label{fig:STC_MC}
\end{figure*}

\subsection{OLMC-SE-UCCD Solution} 
\label{subsec: Stc_Results}

\begin{table}[t]
    \caption{Summary of SE-UCCD, MSC-WCR, and WCR-UCCD solutions.}
    \label{tab:CS1_results}
    \renewcommand{\arraystretch}{1}
    \centering
    \begin{tabular}{l c c c c }
    \hline \hline
    \textrm{\textbf{Formulation}} & $\tilde{o}$ & $\mu_{k}$  & $t (\textrm{s})$ & $t_{switch}$ \\
    \hline
    \textrm{CCD} & $0.301$  & $3.326$ & $1.59$ & $0.727$ \\
    \hline
    \textrm{OLMC-SE-MCS} & $0.304$ & $2.697$  & $5454$ & $0.707$\\
    \textrm{OLMC-SE-gPC} & $0.302$ & $2.537$  & $785$ & $0.716$ \\\cline{2-5}
    $\vert \Delta \vert$ & $0.77\%$ & $5.95\%$  & $85.6\%$ & $1.27\%$\\
    \hline
    {MSC-WCR} & $0.24$ & $3.11$ & $377.9$ & -- \\ 
    \hline 
    \text{OLMC-WCR} & $0.156 $  & $6.4$ & $18.5 $ & $0.75$\\
    \hline \hline
    \end{tabular}
\end{table}

The SE-UCCD problem was solved using an MCS and gPC approach.
Specifically, using $N_{\textrm{mcs}} = 10,000$ number of samples for the MCS-based stochastic UCCD problem, the solution (which is described in Table~\ref{tab:CS1_results}) converged to the objective value of $\tilde{o} = 0.304$, with $\tilde{k}= 2.697$.  
This solution offers a relatively small error of $\mathcal{O}(0.01)$ for the objective function and, thus, is used to benchmark results from gPC.
The solution from gPC differs from that of MCS by $0.77\%$, and $5.95\%$ for $\tilde{o}$ and $\tilde{k}$, respectively.
However, the computational time is reduced by $85.6\%$, which is a considerable improvement over the MCS method. 
Since the open-loop optimal behavior of the system is bang-bang with a single switch, the average control switching time $t_{switch}$ was also estimated when relevant.
From Table~\ref{tab:CS1_results}, the average control switching time for gPC differs from that of MCS by $1.27\%$. 
It is expected that the results from gPC can be further improved by increasing the number of nodes.

Compared to the deterministic CCD solution, the results indicate an increase in the expected value of the objective function.
This increase is justified by the fact that the SE-UCCD is a risk-neutral formulation in which no measure is taken to move away from uncertainties.
Therefore, depending on their distribution, some uncertainty realizations may act in favor of the objective function. 
Therefore, the objective function may witness an increase or a decrease compared to the deterministic solution.  
Note, however, that this is generally not the case for stochastic chance-constrained formulations because their risk-averse nature pushes the solutions toward the interior of the feasible space to maintain the desired probability of success.

Optimal state and control trajectories, along with their associated $\pm\sigma$ distribution bands for MCS and gPC-based implementations, are shown in Fig.~\ref{fig:STC_MC} and compared with the deterministic solution.  
The results indicate that gPC is very well capable of estimating statistics of problem elements.
Specifically, according to Figs.~\ref{fig:STC_MC_MCS_st} and \ref{fig:STC_MC_gPC_st}, the expected value response of relative displacement from both MCS and gPC-based methods are slightly higher than their associated deterministic counterparts.
This observation is also aligned with the increase in the objective function value.
In addition, the OLMC structure allows the optimizer to change the controller's switching time in response to the realization of uncertainties.
This results in a range of optimal control responses whose distribution band with $\pm \sigma_{\tilde{u}}$ is shown in Figs.~{\ref{fig:STC_MC_MCS_ctrl}} and {\ref{fig:STC_MC_gPC_ctrl}} for MCS and gPC, respectively.
Note that while the control is expected to remain bang-bang in nature, the intermediate control quantities in these trajectories seem to be a by-product of the optimization algorithm and/or UP methods and are included here for demonstration purposes.
In addition, note that the average control switching time shows a $2.75\%$ difference between MCS and deterministic CCD solutions.

\subsection{MSC-UCCD Solution}
\label{subsec:MPC_Results}
Representing uncertainties as a set of discrete plants through polytopic descriptions, the UCCD problem of the simple SASA system was also investigated through a multi-stage control approach (similar to the multi-stage robust MPC).
The robust horizon $\Delta t_{r}$ was set to a single time-step using $n_{t} = 100$.

\begin{figure*}[t]
\captionsetup[subfigure]{justification=centering}
\centering
\begin{subfigure}{0.25\textwidth}
\centering
\includegraphics[scale=0.33]{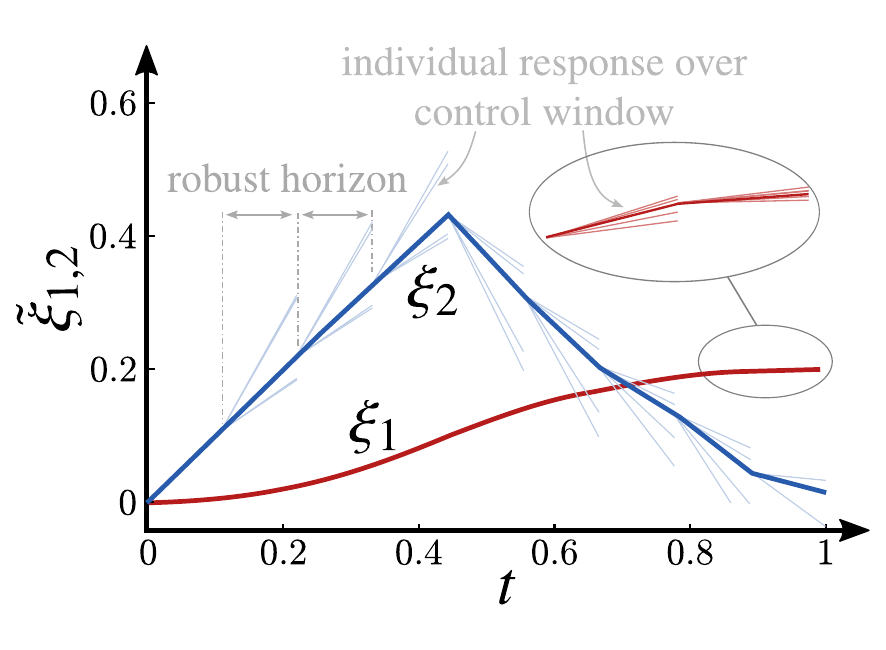}
\caption{Optimal states with $n_{t}=10$.}
\label{fig:MPCR_State10}
\end{subfigure}%
\begin{subfigure}{0.25\textwidth}
\centering
\includegraphics[scale=0.33]{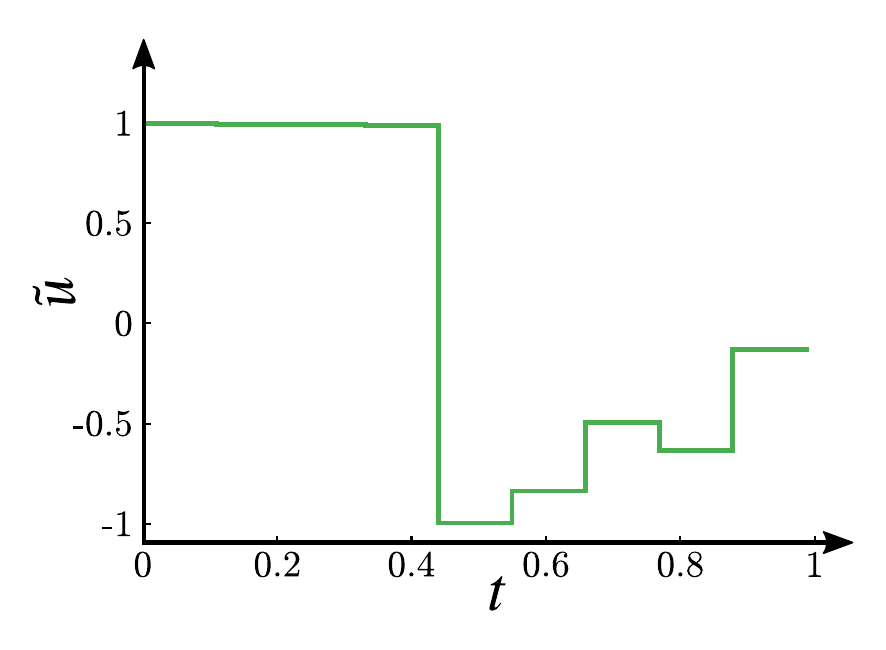}
\caption{Optimal control with $n_{t}  =10$.}
\label{fig:MPCR_Control10}
\end{subfigure}%
\begin{subfigure}{0.25\textwidth}
\centering
\includegraphics[scale=0.33]{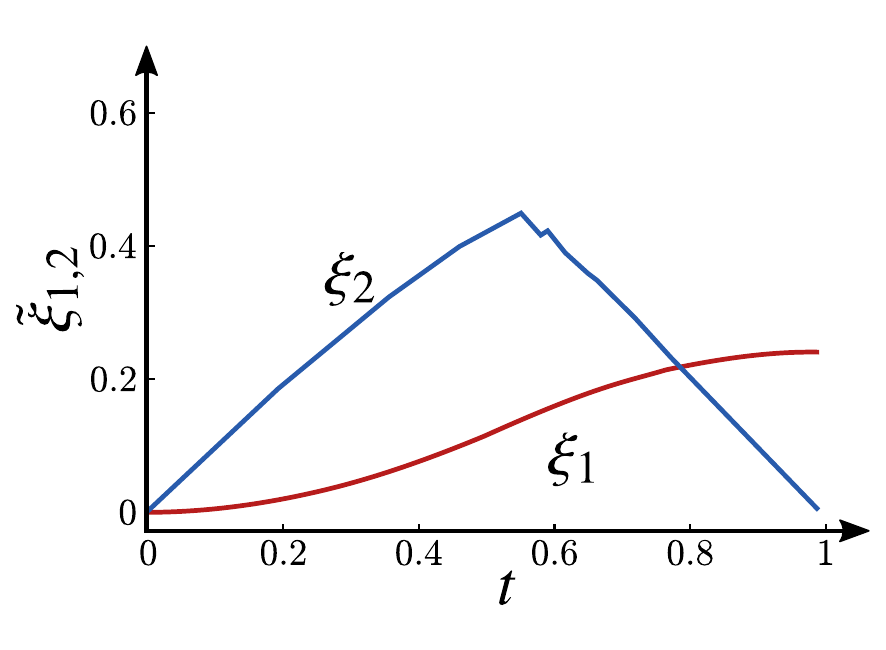}
\caption{Optimal states with $n_{t} = 100$.}
\label{fig:MPCR_State100}
\end{subfigure}%
\begin{subfigure}{0.25\textwidth}
\centering
\includegraphics[scale=0.33]{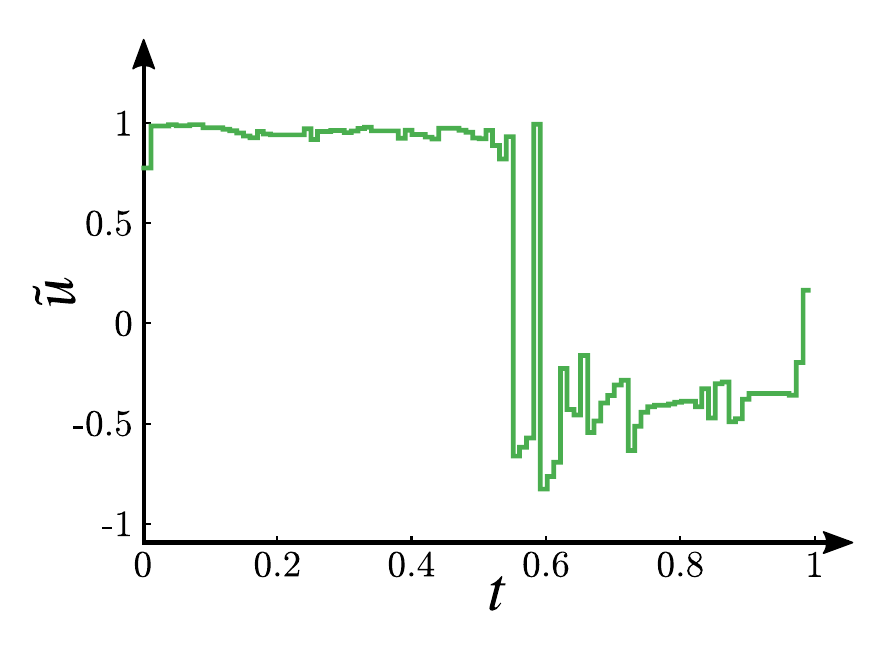}
\caption{Optimal control with $n_{t} = 100$.}
\label{fig:MPCR_Control100}
\end{subfigure}%
\captionsetup[figure]{justification=centering}
\caption{Multi-stage control solution for the simple SASA UCCD problem with $n_{t} = 10$ and $n_{t} = 100$.} 
\label{fig:MPCResults}
\end{figure*}

The multi-stage UCCD solution resulted in $\tilde{o} = 0.24$ and $\tilde{k} = 3.11$, indicating a $63.85\%$ improvement in the performance, and $51.41\%$ drop in $\tilde{k}$ compared to OLMC-WCR-UCCD.
This solution, however, is obtained with an increased computational cost of $377.9~s$. 
The optimal state and control trajectories for two implementations of the multi-stage problem, using $n_{t} = 10$, and $n_{t} = 100$ in each inner-loop subproblem, is shown in Fig.~(\ref{fig:MPCResults}). 
These plots also show the individual response of each plant over the robust horizon.
Note, however, that due to the presence of feedback, the true state of the system is updated at every time step, resulting in robust performance that does not have the conservativeness of the WCR-UCCD solution.  

\begin{figure*}[t]
\captionsetup[subfigure]{justification=centering}
\centering
\begin{subfigure}{0.25\textwidth}
\centering
\includegraphics[scale=0.33]{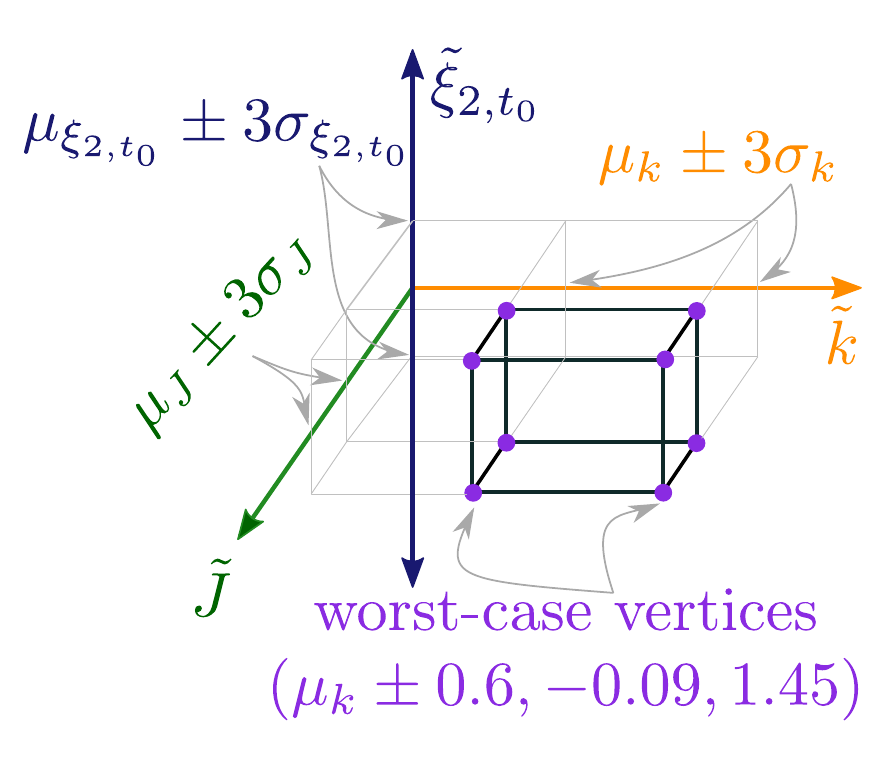}
\caption{Polytope formed by uncertainties.}
\label{fig:WCR_MCpolytope}
\end{subfigure}%
\begin{subfigure}{0.25\textwidth}
\centering
\includegraphics[scale=0.33]{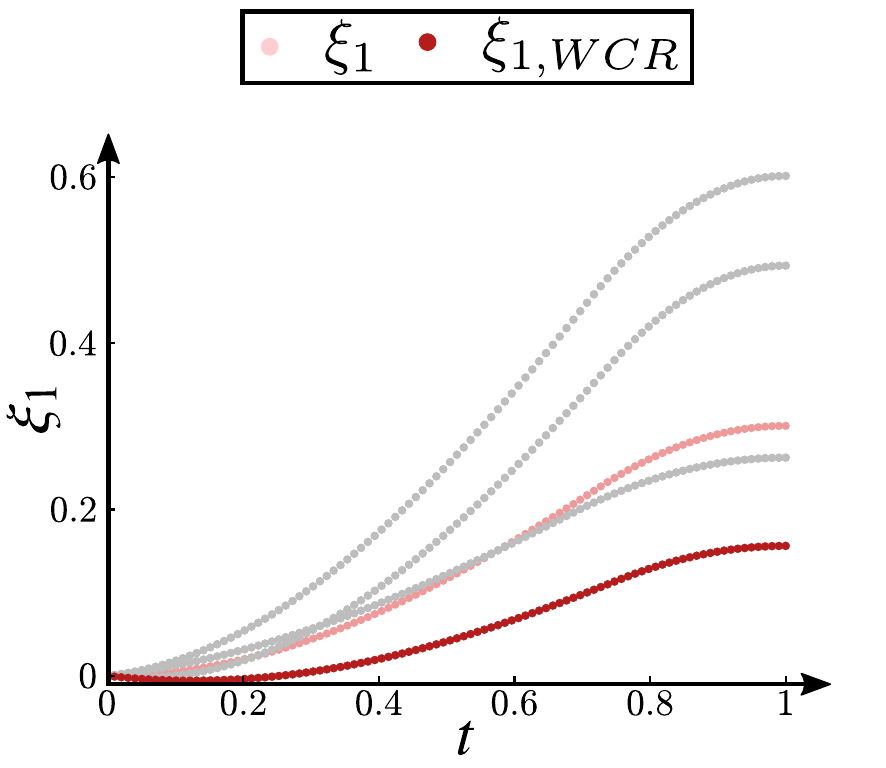}
\caption{$\xi_{1}$ at polytope vertices.}
\label{fig:WCR_MCState1}
\end{subfigure}%
\begin{subfigure}{0.25\textwidth}
\centering
\includegraphics[scale=0.33]{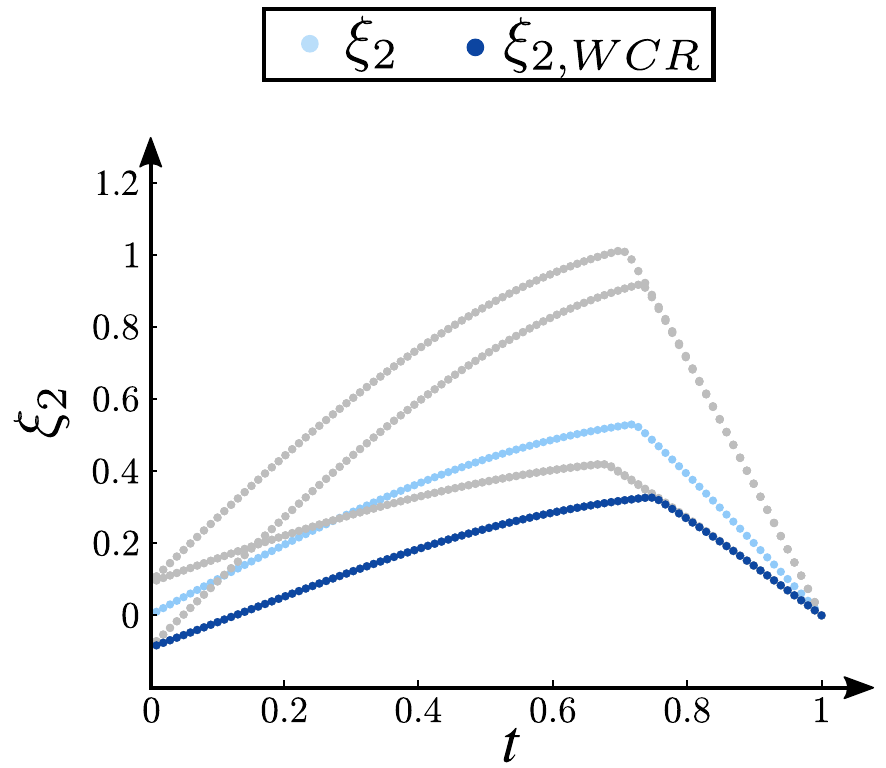}
\caption{$\xi_{2}$ at polytope vertices.}
\label{fig:WCR_MCState2}
\end{subfigure}%
\begin{subfigure}{0.25\textwidth}
\centering
\includegraphics[scale=0.33]{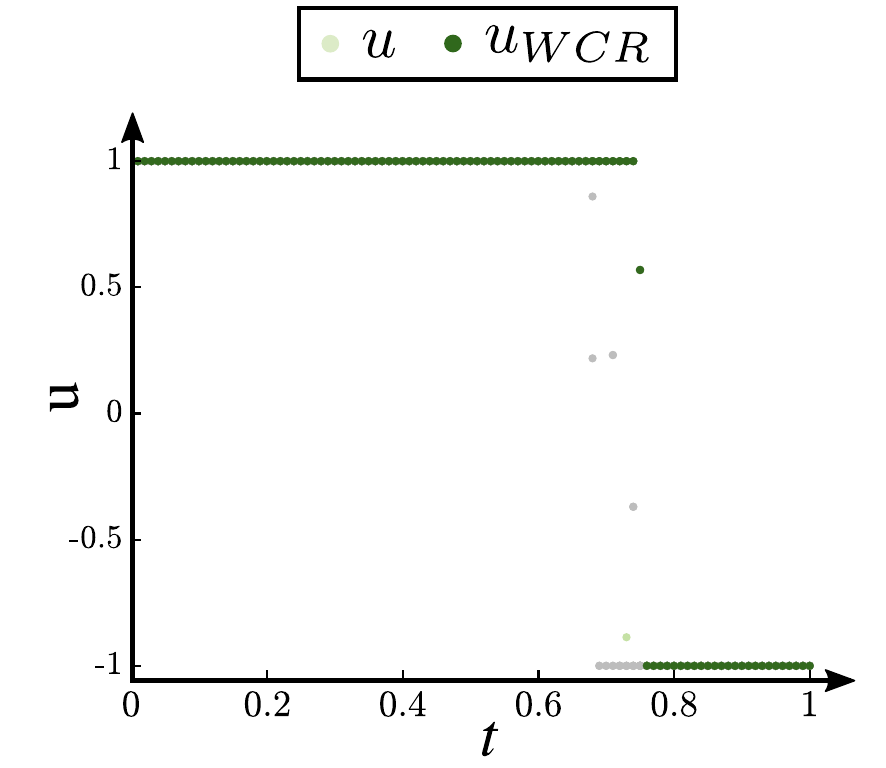}
\caption{$u$ at polytope vertices.}
\label{fig:WCR_MCcontrol}
\end{subfigure}%
\captionsetup[figure]{justification=centering}
\caption{Open-loop multiple-control (OLMC) implementation of the WCR-UCCD for the simple SASA problem using polytopic uncertainties.}
\label{fig:WCR_MC}
\end{figure*}

\subsection{OLMC-WCR-UCCD Solution}
\label{subsec:WcR_Results}

The WCR-UCCD problem is investigated using the concept of polytopic uncertainties, and its solution is obtained by solving Eqs.~(\ref{Eqn:SASA_WCUCCD_MC_O})--(\ref{Eqn:SASA_WCUCCD_MC}) at the vertices of the polytope formed by uncertain quantities.
The results from this implementation are tabulated in Table~\ref{tab:CS1_results}.
Specifically, after solving the WCR-UCCD problem at all of the $2^3$ vertices of the polytope, it turns out that the worst-case realization of uncertainties corresponds to the combination of $(\mu_{k}\pm k_{s}\sigma_{k} = 6.4, -k_{s}\sigma_{\xi_{2,t_{0}}} = -0.09, k_{s}\sigma_{J} = 0.45)$ and results in the objective function of $0.156$, which indicates a $48.17\%$ and $48.68\%$ reduction compared to the deterministic CCD and SE-UCCD (using MCS), respectively. 
The solar array stiffness, on the other hand, has increased by $92.4\%$ and $137.3\%$ compared to the deterministic CCD and SE-UCCD (using MCS) solutions, respectively.
Note that this stiffness value is reported from a vertex in which the worst-case realization for solar array stiffness is described as $\mu_{k}+0.6 = 6.4$.
The higher value of solar array stiffness indicates that for the same amount of displacement, the WCR-UCCD solution would require more force.  
As we will see in Sec.~\ref{subsec: WCR_OLMC_Results}, the high value of stiffness offers some desirable characteristics that enable the system to meet certain requirements, such as quick damping.
The computational time associated with solving the WCR-UCCD at all of the vertices is $18.5$~s, and the switching time associated with the worst realization of uncertainties is $0.75$~s.

The optimal state and control trajectories at the vertices of the polytope and their comparison with the deterministic CCD solution are shown in Fig.~\ref{fig:WCR_MC}. 
Specifically, Fig.~\ref{fig:WCR_MCpolytope} illustrates the polytope formed by uncertainty sets, with a total of $2^3$ vertices.
In these figures, all gray trajectories are associated with vertices that do not represent the worst-case realization of uncertainties. 
Figures \ref{fig:WCR_MCState1}-\ref{fig:WCR_MCcontrol} shows $\xi_{1}$, $\xi_{2}$, and $u$ trajectories for all of the vertices of the polytope.
Note that while the polytope has a total of $8$ vertices, some trajectories completely overlap with others (due to problem symmetry), resulting in a total of $4$ distinct trajectories for states.  
Each pair of overlapping trajectories have $J$ and $\xi_{2,{t_{0}}}$ in common, but differ in $k$ dimension.
Since $\mu_{k}$ is an optimization variable, the optimizer selects this value such that the total stiffness in the optimization problem amounts to the same value \textemdash{} resulting in identical trajectories.

\begin{figure}[t]
\centering
\begin{subfigure}{0.5\columnwidth}
\centering
\includegraphics[scale=0.28]{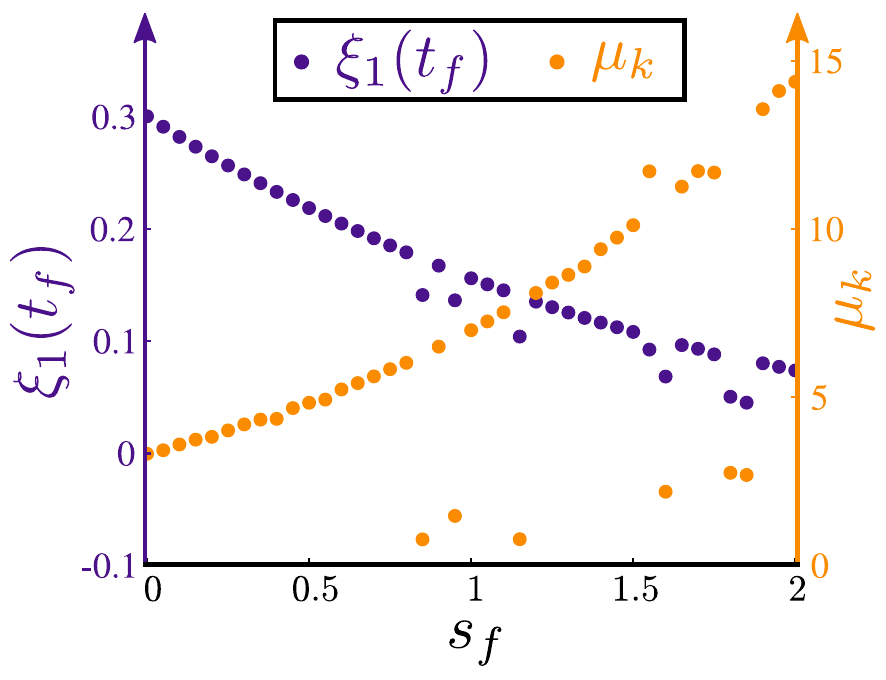}
\caption{WCR solution versus $s_{f}$.}
\label{fig:Uncertaintyscalea}
\end{subfigure}%
\begin{subfigure}{0.5\columnwidth}
\centering
\includegraphics[scale=0.28]{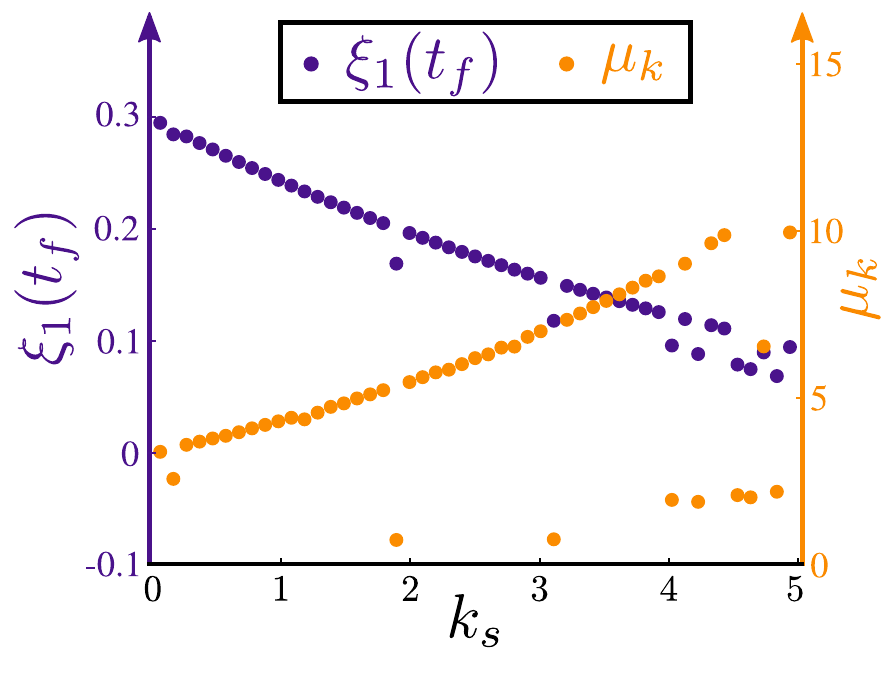}
\caption{WCR solution versus $k_{s}$.}
\label{fig:Uncertaintyscaleb}
\end{subfigure}
\caption{OLMC-WCR-UCCD objective and solar array stiffness as a function of (a) $s_{f}$ which scales standard deviation of uncertainties, and (b) $k_{s}$ which characterizes the size of the uncertainty set.}
\label{fig:Uncertaintyscale}
\end{figure}

It is clear from Fig.~\ref{fig:WCR_MC} that uncertainties in this case study are relatively large because, under the worst realization of uncertainties, the performance of the system is significantly affected by $48.17\%$.
Invaluable insights from this observation can then be offered to the design team or manufacturer.
For example, by investing in higher-precision equipment and measurement devices or improving the precision of the manufacturing processes, the manufacturer may, to some extent, reduce these uncertainties and, therefore, improve the worst-case performance of the system.
To illustrate this, we use a size factor, $s_f$, to scale (up/down) the standard deviation of uncertainties from their current values, such that $\hat{\sigma}_{q} = s_{f} \sigma_{q}$, where $\hat{\sigma}_{q}$ is the new standard deviation.  
Therefore, $s_{f}=1$ corresponds to the current standard deviations, $s_{f} < 1$ corresponds to smaller standard deviations, and $s_{f} > 1$ corresponds to larger standard deviations for uncertainties.

The objective function, $\xi_{1}(t_{f})$ and solar array stiffness, $\mu_{k}$ for different $s_{f}$ values are shown in Fig.~\ref{fig:Uncertaintyscalea}.
From this figure, we can see that as uncertainties increase, the system's performance drops below $0.1$. 
In addition to the standard deviation of uncertainties, the size of the crisp uncertainty set is characterized through $k_{s}$.
As opposed to $s_{f}$, which represents the inherent magnitude of uncertainties, $k_{s}$ reflects designers' understanding of the size of the uncertainty set.
In other words, $k_{s}$ is ideally selected such that the resulting uncertainty set is a close approximation of uncertainty realizations.
If $k_{s}$ is too small, the resulting uncertainty set will not be a good representation, and thus, the WCR-UCCD solution will not be able to protect the system against risks associated with uncertainties.
On the other hand, when $k_{s}$ is too large, the solution will be over-conservative, and thus, the system may be subject to low performance due to over-design.
To investigate this issue, Fig.~\ref{fig:Uncertaintyscaleb} shows the objective function and solar array stiffness values as functions of $k_{s}$ when $s_{f} = 1$ (i.e.,~current standard deviation of uncertainties).
Note that, in these figures, there are some points that do not completely follow the general trend in the data. 
This outcome seems to be due to the optimization solver in the outer loop of the nested coordination strategy.

\subsection{Closed-loop Investigations} 
\label{subsec: WCR_OLMC_Results}

\begin{figure*}[t]
\captionsetup[subfigure]{justification=centering}
\centering
\begin{subfigure}{0.25\textwidth}
\centering
\includegraphics[scale=0.3]{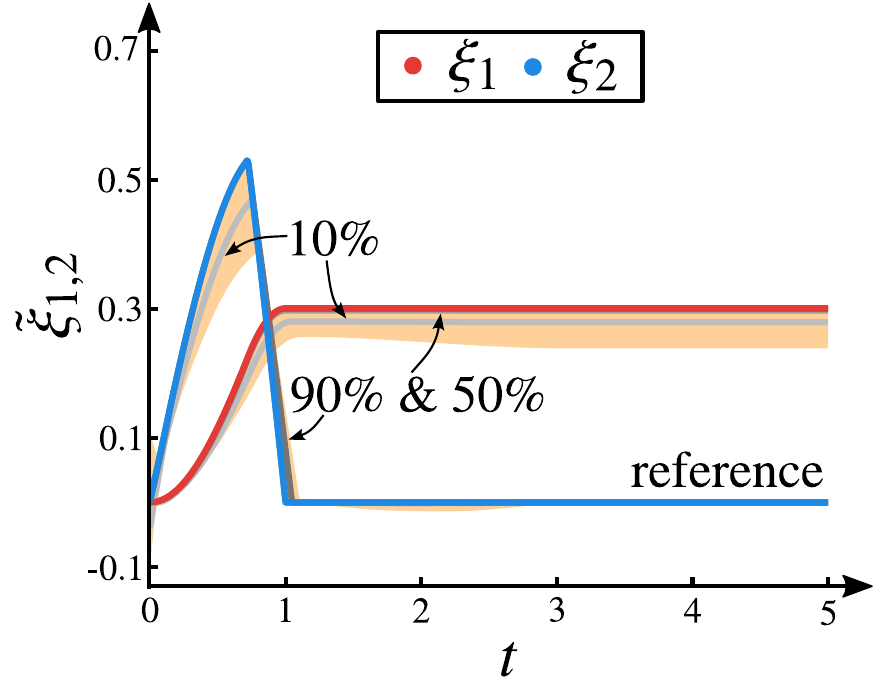}
\caption{DET-SYS performance.}
\label{fig:CL_Det_sP}
\end{subfigure}%
\begin{subfigure}{0.25\textwidth}
\centering
\includegraphics[scale=0.3]{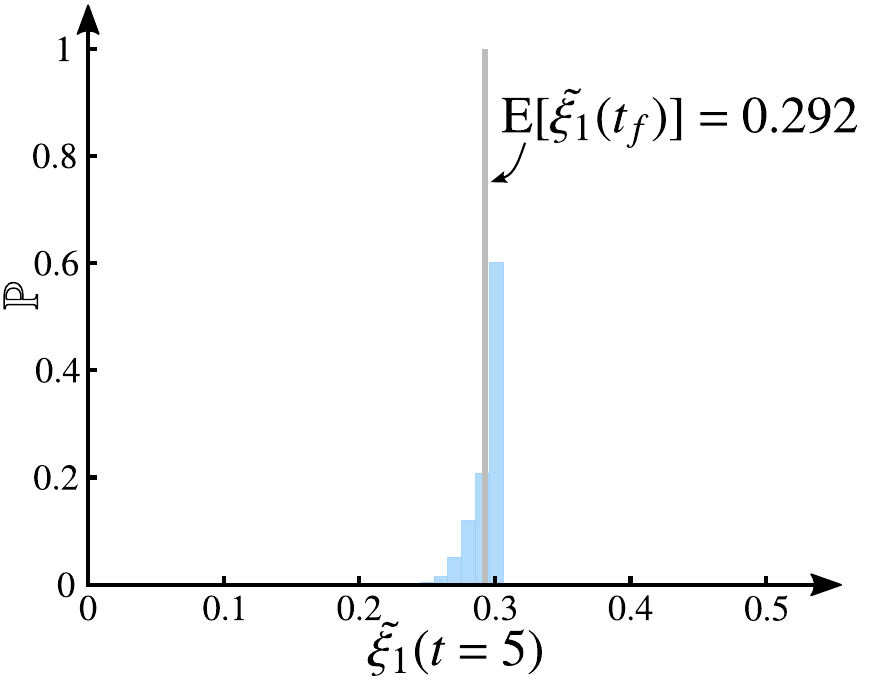}
\caption{DET-SYS $\tilde{\xi_{1}}(t_f)$ distribution.}
\label{fig:CL_Det_oP}
\end{subfigure}%
\begin{subfigure}{0.25\textwidth}
\centering
\includegraphics[scale=0.3]{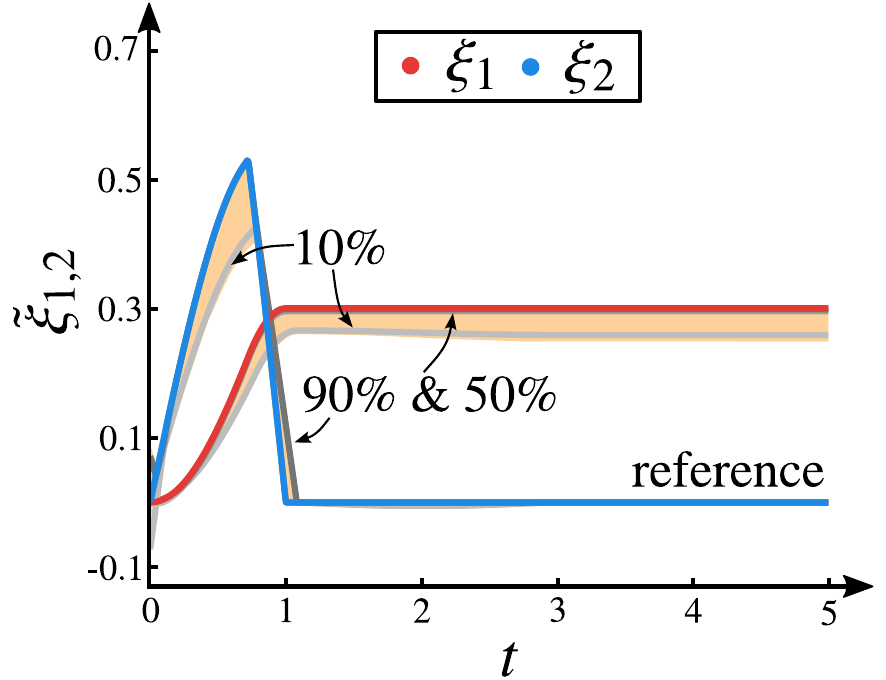}
\caption{DET-SYS performance.}
\label{fig:CL_Det_sB}
\end{subfigure}%
\begin{subfigure}{0.25\textwidth}
\centering
\includegraphics[scale=0.3]{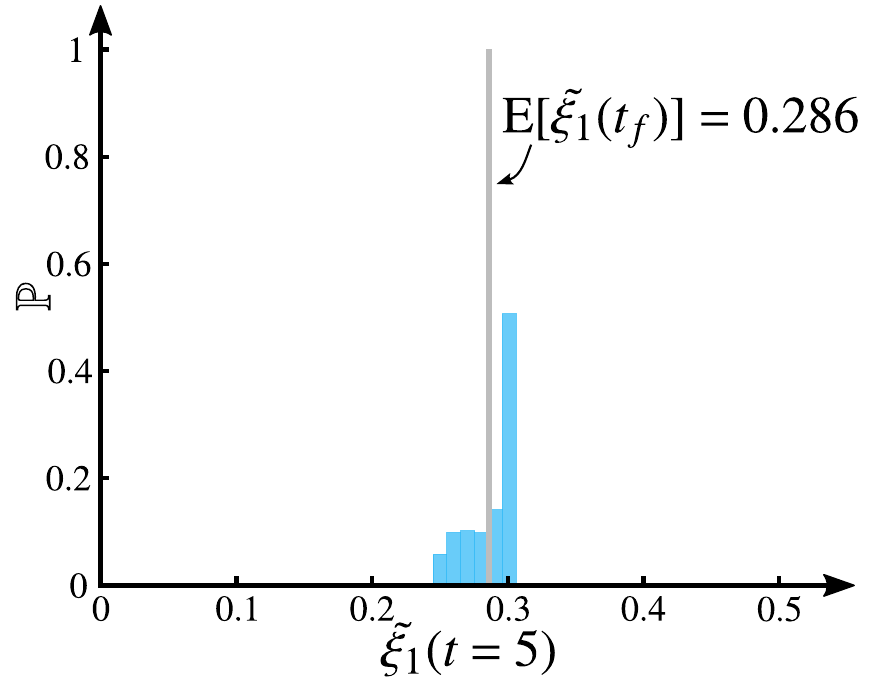}
\caption{DET-SYS $\tilde{\xi_{1}}(t_f)$ distribution.}
\label{fig:CL_Det_oB}
\end{subfigure}%
\captionsetup[figure]{justification=centering}
\caption{DET-SYS performance with Gaussian samples in (a) and (b) and samples from a uniform distribution in (c) and (d).} 
\label{fig:Cl_DET}
\end{figure*}

\begin{figure*}[t]
\captionsetup[subfigure]{justification=centering}
\centering
\begin{subfigure}{0.25\textwidth}
\centering
\includegraphics[scale=0.3]{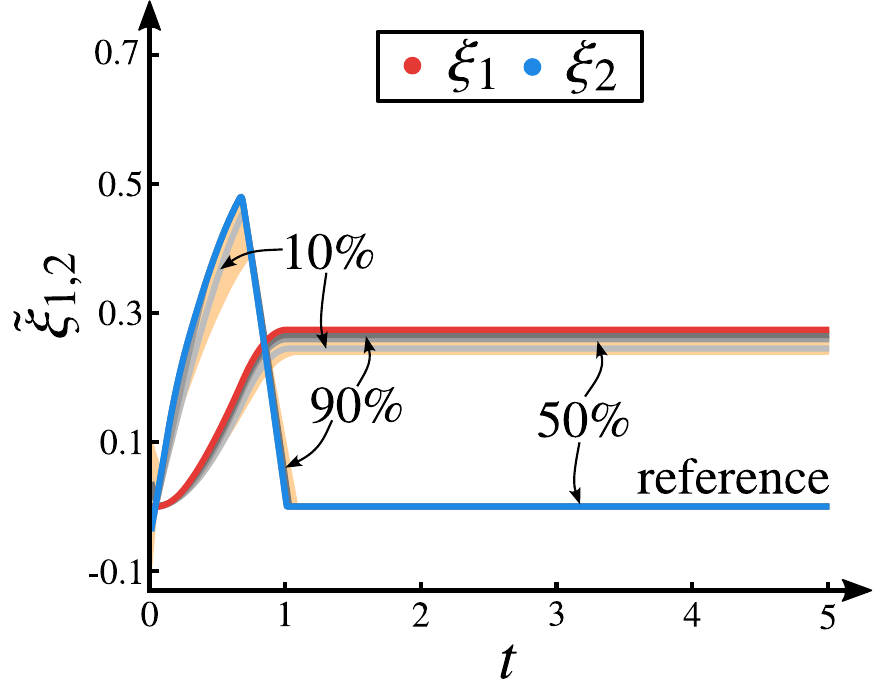}
\caption{SE-SYS-10 performance.}
\label{fig:CL_SE_s10}
\end{subfigure}%
\begin{subfigure}{0.25\textwidth}
\centering
\includegraphics[scale=0.3]{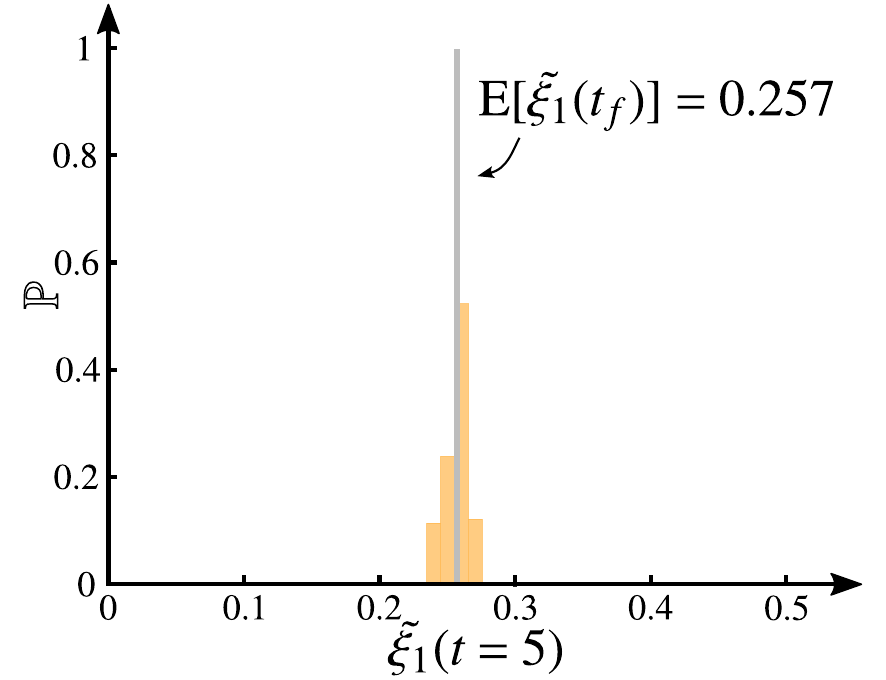}
\caption{SE-SYS-10 $\tilde{\xi_{1}}(t_f)$ distribution.}
\label{fig:CL_SE_o10}
\end{subfigure}%
\begin{subfigure}{0.25\textwidth}
\centering
\includegraphics[scale=0.3]{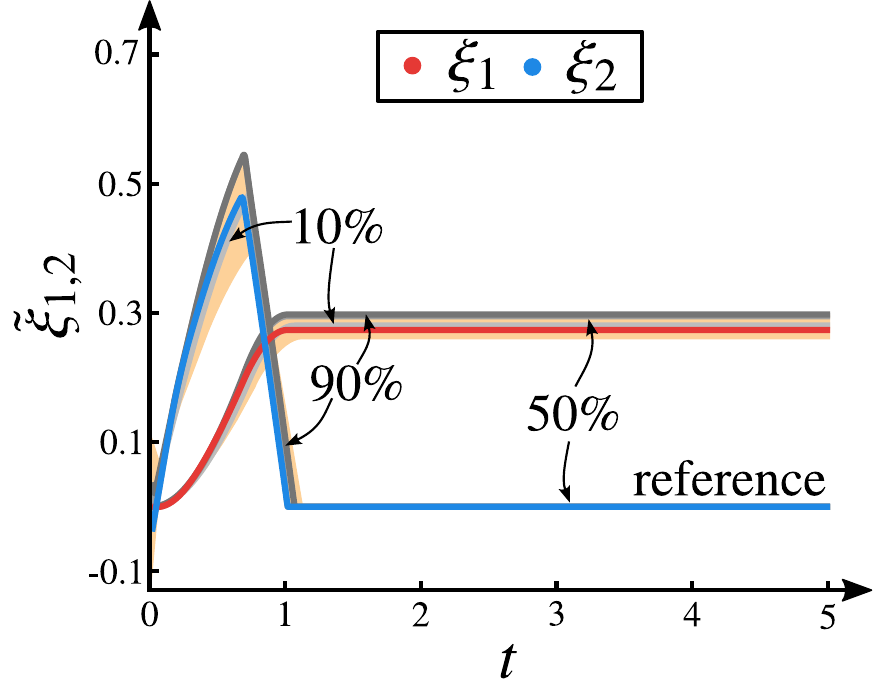}
\caption{SE-SYS-50 performance.}
\label{fig:CL_SE_s50}
\end{subfigure}%
\begin{subfigure}{0.25\textwidth}
\centering
\includegraphics[scale=0.3]{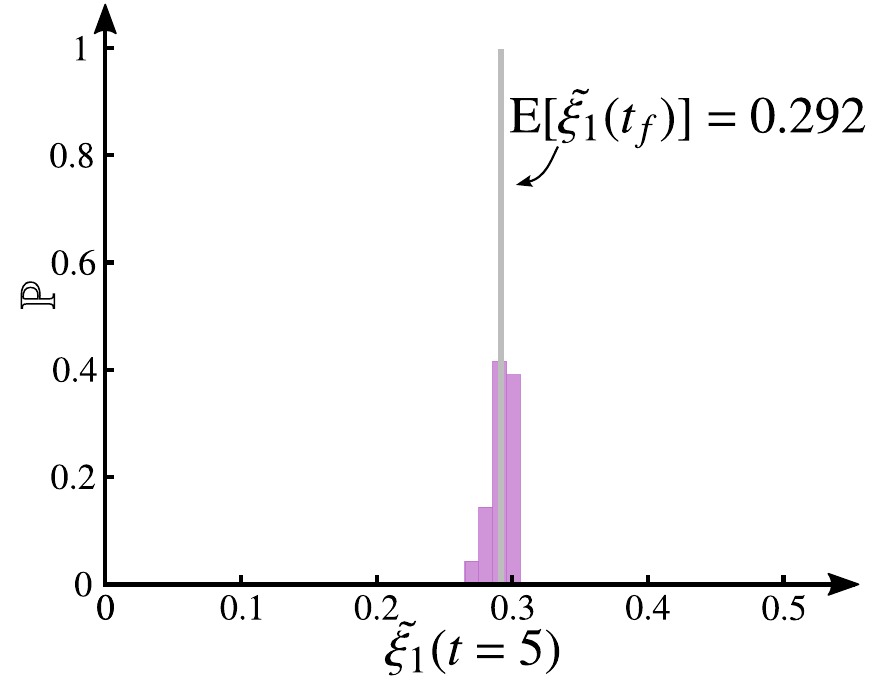}
\caption{SE-SYS-50 $\tilde{\xi_{1}}(t_f)$ distribution.}
\label{fig:CL_SE_o50}
\end{subfigure}
\begin{subfigure}{0.25\textwidth}
\centering
\includegraphics[scale=0.3]{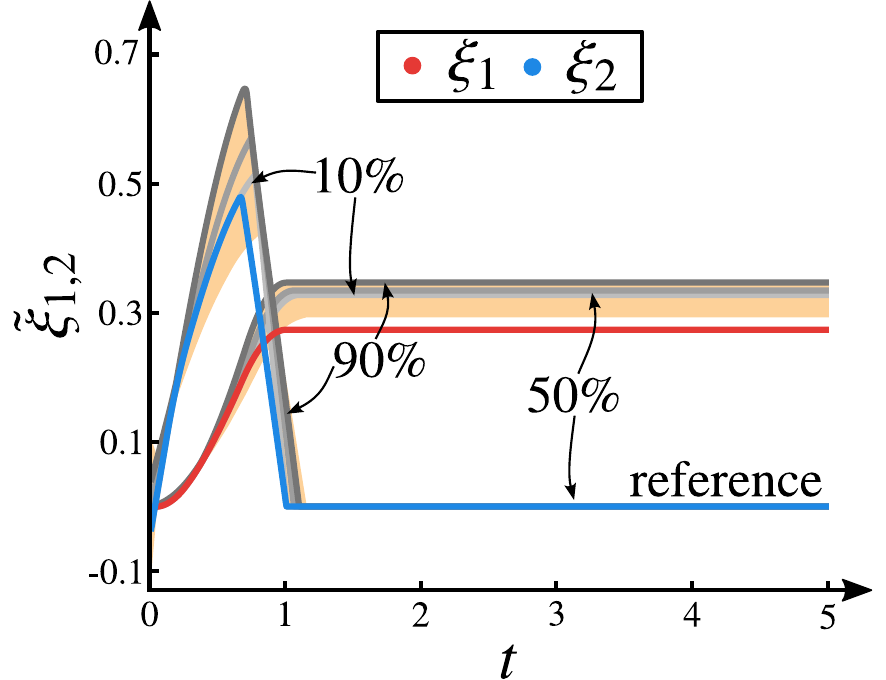}
\caption{SE-SYS-90 performance.}
\label{fig:CL_SE_s90}
\end{subfigure}%
\begin{subfigure}{0.25\textwidth}
\centering
\includegraphics[scale=0.3]{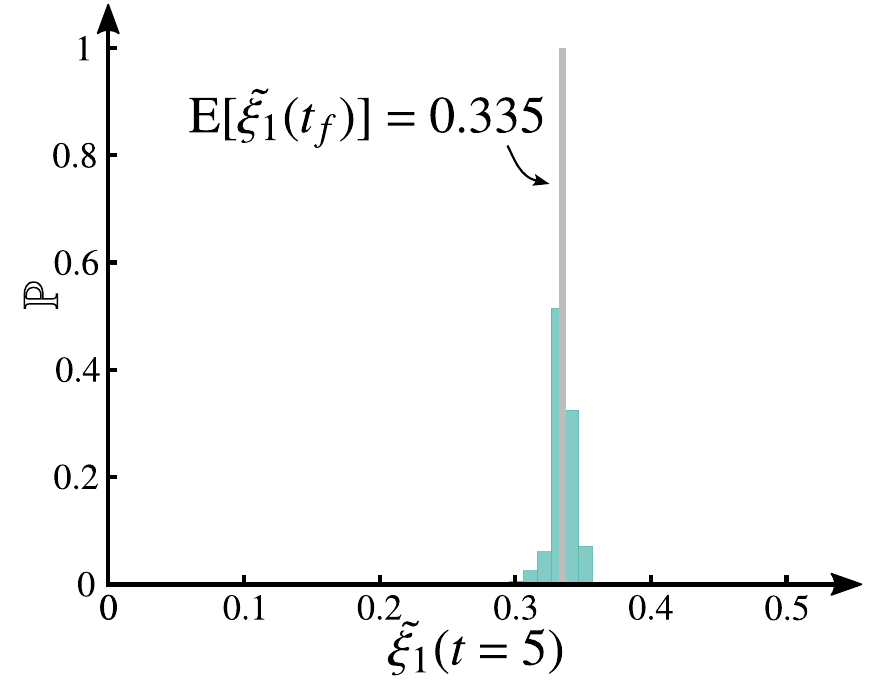}
\caption{DET-SYS-90 $\tilde{\xi_{1}}(t_f)$ distribution.}
\label{fig:CL_SE_o90}
\end{subfigure}%
\begin{subfigure}{0.25\textwidth}
\centering
\includegraphics[scale=0.3]{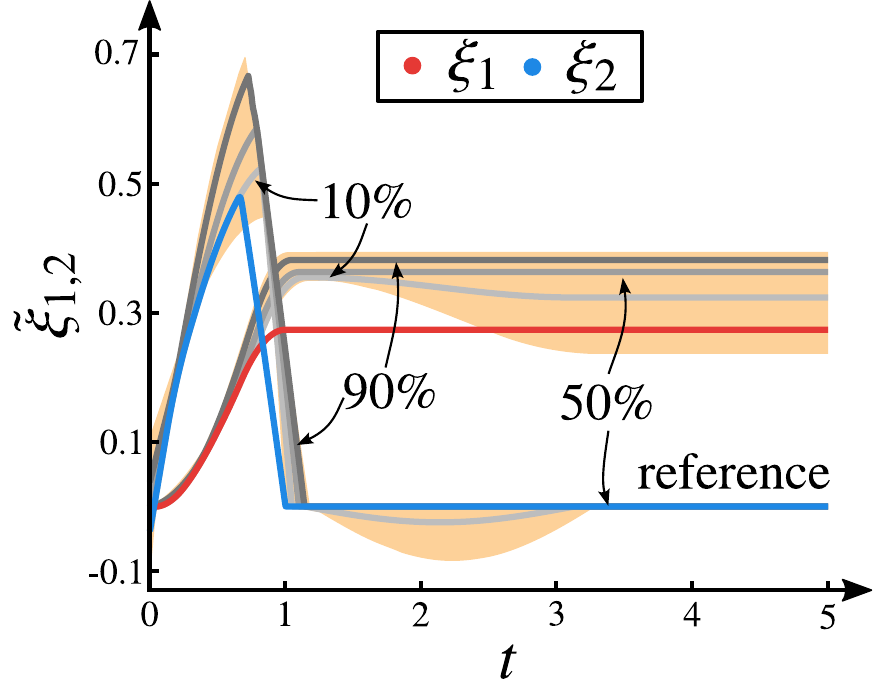}
\caption{SE-SYS-100 performance.}
\label{fig:CL_SE_s100}
\end{subfigure}%
\begin{subfigure}{0.25\textwidth}
\centering
\includegraphics[scale=0.3]{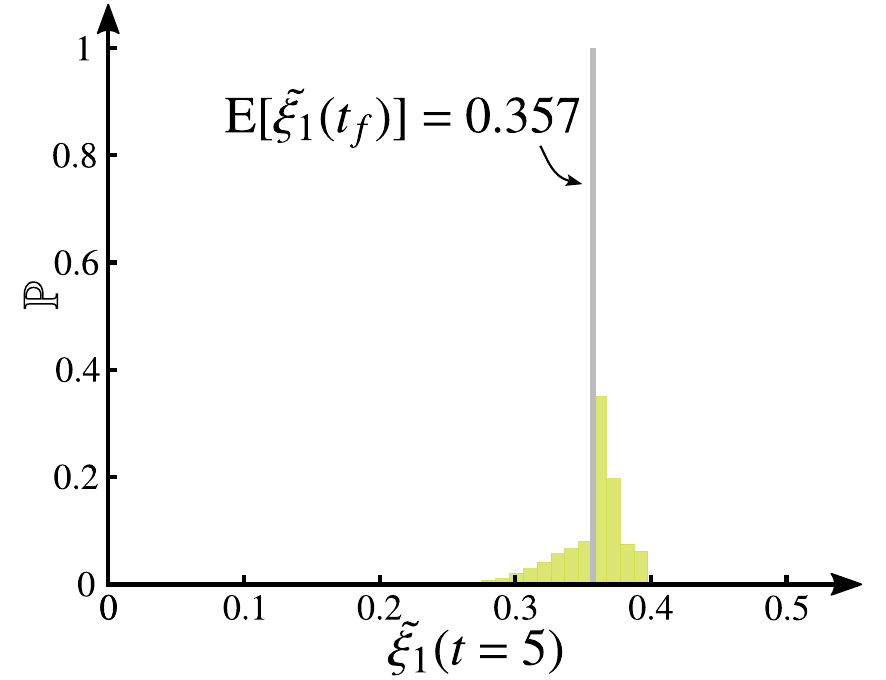}
\caption{SE-SYS-100 $\tilde{\xi_{1}}(t_f)$ distribution.}
\label{fig:CL_SE_o100}
\end{subfigure}%
\captionsetup[figure]{justification=centering}
\caption{SE-SYS performance with Gaussian samples and $10$th, $50$th, $90$th, and $100$th percentiles of optimal trajectories as references.}
\label{fig:Cl_SE}
\end{figure*}

\begin{figure}[t]
\captionsetup[subfigure]{justification=centering}
\centering
\begin{subfigure}{0.5\columnwidth}
\centering
\includegraphics[scale=0.3]{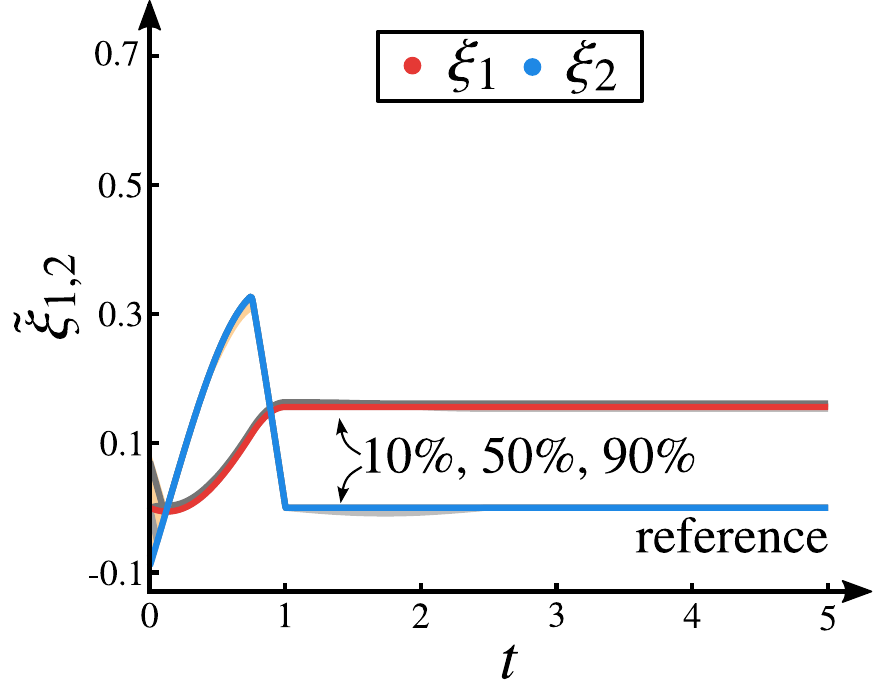}
\caption{WCR-SYS performance.}
\label{fig:CL_WCR_s}
\end{subfigure}%
\begin{subfigure}{0.5\columnwidth}
\centering
\includegraphics[scale=0.3]{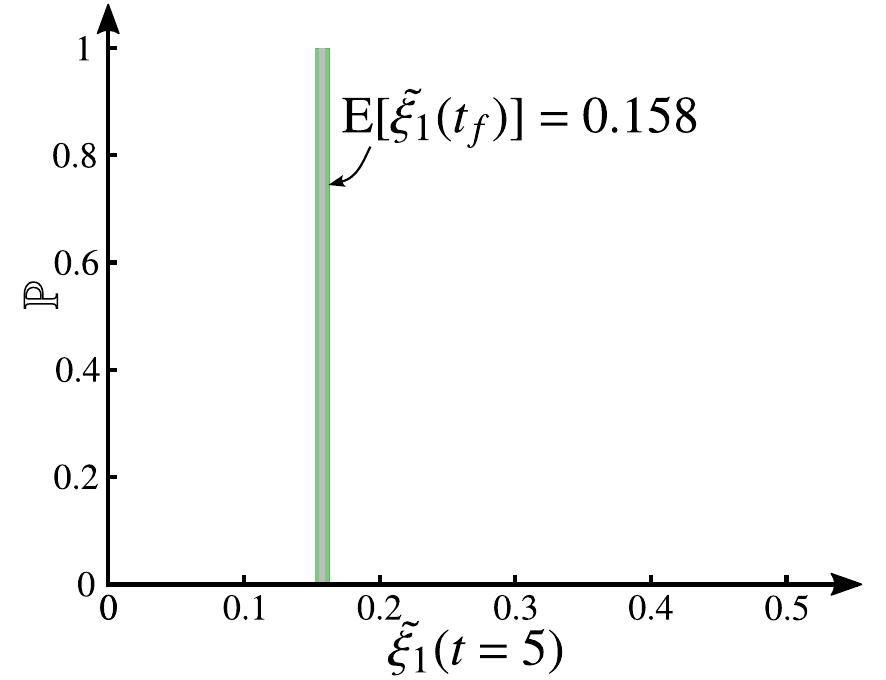}
\caption{WCR-SYS $\tilde{\xi_{1}}(t_f)$ distribution.}
\label{fig:CL_WCR_o}
\end{subfigure}
\captionsetup[figure]{justification=centering}
\caption{WCR-SYS performance with samples from a uniform distribution.}
\label{fig:Cl_WCR}
\end{figure}

\begin{table}[t]
    \caption{Mean and standard deviation of the performance of various closed-loop systems for the simulation time of $t_{f} = 5~\unit{s}$.}
    \label{tab:Closed_loop_stats}
    \renewcommand{\arraystretch}{1}
    \centering
    \begin{tabular}{l l r r r }
    \hline \hline
    \textrm{\textbf{System}} & Samples & $\mu_{\xi_{1}(t_{f})}$ & $ \sigma_{\xi_{1}(t_{f})}$ & $t_{settle}$ \\
    \hline
    {\text{DET-SYS}} & {\text{Stochastic}} & {$0.293$} & {$0.0095$} & {$1.091$} \\
    {\text{DET-SYS}} & {\text{{Crisp}}} & {$0.286$} & {$0.0155$}  & {$1.076$} \\
    {\text{SE-SYS-10}} & {\text{Stochastic}}  & {$0.257$} & {$0.0080$} & {$1.066$}\\
    {\text{SE-SYS-50}} & {\text{Stochastic}}  & {$0.292$} & {$0.0073$} & {$1.096$}\\
    {\text{SE-SYS-90}} & {\text{Stochastic}}  & {$0.335$} & {$0.0082$} & {$1.131$}\\
    {\text{SE-SYS-100}} & {\text{Stochastic}} & {$0.357$} & {$0.0236$} & {$3.128$}\\
    {\text{WCR-SYS}} & {\text{{Crisp}}}   & {$0.158$} & {$0.0025$} & {$1.000$} \\
    \hline \hline
    \end{tabular}
\end{table}

In the next step, design specifications corresponding to deterministic CCD, SE-UCCD (using MCS), and WCR-UCCD solutions are used to synthesize a closed-loop proportional feedback controller for the simple SASA system.
To avoid repetition and to distinguish the following simulations from their corresponding deterministic and/or uncertain CCD optimization problems, we refer to these systems as DET-SYS, SE-SYS, and WCR-SYS, respectively.
Note that while more advanced control strategies might offer improved performance, the simplicity of this closed-loop control architecture allows us to investigate the impact of these formulations on the performance of the system and some of its key characteristics, such as settling time.    
The $10$th, $50$th, $90$th, and $100$th percentiles of optimal control and the second state trajectories are used in the SE-SYS to create the associated references.
These percentiles are augmented to the name, so SE-SYS-10 is the system designed using the SE-UCCD solution, following reference trajectories created from the $10$th percentile of optimal control and state trajectories.

To assess the closed-loop performance of the system, the control gain is selected to aggressively favor reference-tracking over disturbance rejection.
The response of the system is then obtained by creating $10,000$ samples from the uncertain space.
To create samples for crisp uncertainties, a uniform distribution within $\bm{\mu}_{q} \pm k_{s}\sigma_{q}$ was used.
The simulations were implemented using the \textsc{Matlab}-based $ode15s$ solver with relative tolerance of $10^{-13}$ and absolute tolerance of $10^{-14}$, and a simulation time of $t_{f} = 5~$s.
The results from this analysis are tabulated in Table~\ref{tab:Closed_loop_stats} and shown in Figs.~\ref{fig:Cl_DET}-\ref{fig:Cl_WCR}.
These figures also entail trajectories that represent the $10$th, $50$th, and $90$th percentiles of the performance, as well as histogram representations of the distribution of $\tilde{\xi}_{1}(t_{f} = 5)$, normalized to indicate its probability.

Table~\ref{tab:Closed_loop_stats} reports three results from this analysis that are key in understanding systems' behavior in the presence of uncertainties.  
Specifically, $\mu_{\xi_{1}(t_f)}$ and $\sigma_{\xi_{1}(t_f)}$ are the expected value and standard deviation of the terminal displacement for the associated simple SASA system, while $t_{settle}$ is the settling time.
Note that $t_{settle}$ is calculated such that the system is considered to be at rest if the relative velocity is within $\pm 0.02\%$.

From this table, it is noticeable that in the CCD-SYS, the expected displacement is $0.293$ and $0.286$ for stochastic and crisp samples, respectively.
The relatively high standard deviation of relative displacement at terminal time in these cases points to the fact that, under uncertainty realizations, the system's performance will be significantly affected. 
In terms of the settling time, these system go to rest at $1.091~\unit{s}$ and $1.076~\unit{s}$, respectively.
The performance of the system and the distribution of $\tilde{\xi}_{1}(t_f)$ are shown in Figs.\ref{fig:Cl_DET}, in which Figs.~\ref{fig:CL_Det_sP} and \ref{fig:CL_Det_oP} show systems performance with samples from a Gaussian distribution, while Figs.~\ref{fig:CL_Det_sB} and \ref{fig:CL_Det_oB} show systems performance with crisp uncertainties.

According to Table~\ref{tab:Closed_loop_stats}, the performance of the SE-SYSs increases as we introduce a higher percentile of the optimal trajectories as references. 
The SE-SYS-10 and SE-SYS-100 exhibit the lowest and highest performances with $0.257$ and $0.357$, respectively.
This indicates a $38.9\%$ improvement in SE-SYS-100 when compared to SE-SYS-10. 
However, the standard deviation and settling time associated with SE-SYS-100 also increase by $195\%$, and $193\%$, respectively.
This observation signifies some of the trade-offs between the system's performance, its associated dispersion, and the settling time. 
The SE-SYS-50 and SE-SYS-90 represent systems that offer a reasonable trade-off between these key criteria.
Among these, the SE-SYS-90 offers a relatively high performance of $0.335$ with a reasonable settling time of $1.131~\unit{s}$ and a standard deviation of $0.0082$ \textemdash{} making it a good candidate for designs that emphasize a balance between these criteria.  
The analysis associated with these systems is presented in Fig.~\ref{fig:Cl_SE}.
From this figure, it is clear that the distribution band associated with uncertainty realization is much higher in the SE-SYS-100. 
It is also clear that the settling time is much worse compared to other systems.

According to Table \ref{tab:Closed_loop_stats}, the WCR-SYS offers the lowest performance with the lowest standard deviation and settling time.
Specifically, the performance of the WCR-SYS has decreased by $44.76\%$ compared to the DET-SYS with crisp uncertainties. 
This is directly associated with the conservativeness of the WCR-UCCD approach.
The WCR-SYS analyses also present the least standard deviation, as well as the settling time among the systems studied.
These observations can be visually confirmed through Figs.~\ref{fig:CL_WCR_s} and \ref{fig:CL_WCR_o}.

Overall, the analysis sheds some light on the behavior of each of these systems.
It highlights the fact that different design criteria might favor different systems.
For example, when maximum performance is the principle design criterion, then the SE-SYS-100 might be the best candidate.
Note, however, that while offering the highest expectation, the performance of the SE-SYS-100 can deviate significantly from its expected performance. 
The SE-SYS-90 offers the best compromise between performance, settling time, and objective dispersion. 
The WCR-SYS, on the other hand, is conservative, but offers a consistence performance when uncertainties are realized.  
The robustness of this system is evident from Figs.~\ref{fig:CL_WCR_s}-\ref{fig:CL_WCR_o}, where the distribution band is extremely small, and the system performs within an expected interval with the probability of (almost) one. 
While conservative, the WCR-SYS implemented here overlooks rare events (i.e.,~realizations with probabilities less than $0.3\%$ corresponding to samples beyond $\pm 3\sigma$) in the design and analysis of the system.

These investigations clearly show that uncertainties should be treated as a critical element in dynamic system design from the early stages and not as an afterthought. 
In addition, exploration of the whole design space (both plant and control) at the early stages of the design process, through OLMC structures and appropriate UCCD formulations, along with appropriate uncertainty considerations, enables designers to understand the general trends and interactions between various system elements and have reasonable expectations of the system performance under uncertainties.

\xsection{Conclusion}
\label{sec:conclusion}
Implementation of uncertain control co-design (UCCD) formulations for challenging dynamic systems requires an improved understanding of the characteristics of their corresponding solutions.    
For a given application, the availability of information for uncertainty representations might be the primary factor in selecting the appropriate UCCD formulations.
However, if the characteristics of UCCD solutions are considered, it might become evident that the application in hand might benefit from a specific formulation.
Therefore, insights into the characteristics of UCCD solutions may play a central role in selecting the appropriate UCCD formulation for different applications.

In this study, we started by introducing the open-loop multiple control (OLMC), multi-stage control (MSC), and open-loop single control (OLSC) structures for UCCD problems.
Regardless of the choice of control structure, propagation of probabilistic uncertainties in the dynamic system requires efficient uncertainty propagation methods. 
Therefore, we also described and implemented Monte Carlo Simulation (MCS) and generalized Polynomial Chaos (gPC) expansion for stochastic in expectation UCCD problems.
The worst-case robust UCCD (WCR-UCCD) formulation benefited from the linearity of the inner-loop optimal control problem in the nested coordination strategy and polytopic uncertainties.
To address the conservativeness of the WCR-UCCD solution, a multi-stage control structure, motivated by concepts from robust multi-stage model predictive control, was utilized.  
Finally, the closed-loop reference-tracking response of the resulting systems provided additional insights into the characteristics of each of the UCCD solutions.

A natural next step is to extend this work to investigate problems with probabilistic path constraints, with an emphasis on stochastic chance-constraints UCCD formulations.
The inclusion of time-dependent disturbances in the dynamic system model, along with its uncertainty propagation through an efficient and effective method, is a crucial step in implementing UCCD formulations for real-world applications.
The implementation of the Galerkin type of gPC, which is promising for problems with a small number of state variables, might provide an improvement over the collocation type of gPC for small-scale problems.
Comparison with the most-probable-point (MPP)-based UP methods, which have been already studied in reliability-based CCD, may offer additional insights into possible improvements in computational time, especially because the number of dynamic system equations can be reduced significantly.
For WCR-UCCD, various geometries (such as ellipsoidal, hexagonal, etc.) for the uncertainty set must be considered.
Since empirical information about uncertainties is generally limited in early-stage design, it is necessary to investigate non-probabilistic propagation methods such as interval analysis and methods from fuzzy programming.
Finally, to address the question of scalability, the implementations of such methods and formulations for larger problems must be investigated.

\begin{acknowledgment}
This research was partially supported by the National Science Foundation, Division of Civil, Mechanical, \& Manufacturing Innovation, Engineering Design and System Engineering Program, under grant number CMMI-2034040.
\end{acknowledgment}

\renewcommand{\refname}{REFERENCES}
\bibliographystyle{asmems4}
{\small
\bibliography{References}
}


\clearpage


\nomenclature[A, 01]{AAO}{all-at-once}
\nomenclature[A, 02]{a.s.}{almost surely}
\nomenclature[A, 03]{CCD}{control co-design}
\nomenclature[A, 04]{CTR}{composite trapezoidal rule}
\nomenclature[A, 05]{DSS}{direct single shooting}
\nomenclature[A, 06]{DT}{direct transcription}
\nomenclature[A, 07]{DTQP}{ direct transcription quadratic programming}
\nomenclature[A, 08]{ED}{equidistant (nodes)}
\nomenclature[A, 9]{gPC}{generalized polynomial chaos}
\nomenclature[A, 10]{MCS}{Monte Carlo simulation}
\nomenclature[A, 11]{MPP}{most-probable-point}
\nomenclature[A, 12]{ODE}{ordinary differential equations}
\nomenclature[A, 13]{OLMC}{open-loop multiple control}
\nomenclature[A, 14]{MPC}{model predictive control}
\nomenclature[A, 15]{MSC}{multi-stage control}
\nomenclature[A, 16]{OLSC}{open-loop single control}
\nomenclature[A, 17]{PDF}{probability distribution function}
\nomenclature[A, 18]{PR}{probabilistic robust}
\nomenclature[A, 19]{SASA}{strain-actuated solar array}
\nomenclature[A, 20]{SE}{stochastic in expectation}
\nomenclature[A, 21]{TR}{trapezoidal rule}
\nomenclature[A, 22]{UCCD}{uncertain control co-design}
\nomenclature[A, 23]{UP}{uncertainty propagation}
\nomenclature[A, 24]{WCR}{worst-case robust}


\nomenclature[v, 01]{\( \pd\)}{vector of problem data}
\nomenclature[v, 02]{\( \mathbb{E}[\cdot]\)}{expected value operator}
\nomenclature[v, 03]{\( F_{\tilde{\bm{x}}}(\bm{x})\)}{joint probability distribution}
\nomenclature[v, 04]{\( \bm{f}(\cdot)\)}{state transition or derivative function}
\nomenclature[v, 05]{\( \bm{g}(\cdot)\)}{inequality constraint vector}
\nomenclature[v, 06]{\( \bm{h}(\cdot)\)}{equality constraint vector}
\nomenclature[v, 07]{\( J\)}{inertia ratio between the solar array and the bus}
\nomenclature[v, 8]{\( k_{s}\)}{constraint shift index}
\nomenclature[v, 9]{\( \ell(\cdot)\)}{Lagrange term}
\nomenclature[v, 10]{\( k\)}{solar array stiffness}
\nomenclature[v, 11]{\( M\)}{dimension of the $n_{x}$-variate polynomials}
\nomenclature[v, 12]{\( m(\cdot)\)}{ Mayer term}
\nomenclature[v, 13]{\( \mathcal{N}(\cdot) \)}{ normal distribution}
\nomenclature[v, 14]{\( N_{mcs}\)}{number of samples in MCS}
\nomenclature[v, 15]{\( \mathbb{N}_{0}^{n_{x}}\)}{set of $n_{x}$-dimensional natural numbers with zero}
\nomenclature[v, 16]{\( n_{d}\)}{number of elements in the vector of problem data}
\nomenclature[v, 17]{\( n_{p}\)}{number of time-independent optimization variables}
\nomenclature[v, 18]{\( n_{s}\)}{number of state optimization variables}
\nomenclature[v, 19]{\( n_{u}\)}{number of control optimization variables}
\nomenclature[v, 20]{\( n_{x}\)}{number of uncertain basic quantities}
\nomenclature[v, 21]{\( o(\cdot)\)}{ objective function}
\nomenclature[v, 22]{\( PC \)}{polynomial chaos degree}
\nomenclature[v, 23]{\( \bm{p}\)}{time-independent optimization variable vector}
\nomenclature[v, 24]{\( \bm{p}_{c}\)}{control gain vector}
\nomenclature[v, 25]{\( \bm{p}_{p}\)}{plant optimization variables}
\nomenclature[v, 26]{\(Q\)}{ total number of collocation nodes}
\nomenclature[v, 27]{\( \tilde{\bm{q}}\)}{column vector of uncertain quantities}
\nomenclature[v, 28]{\( \hat{\bm{q}}\)}{nominal quantities used to construct uncertainty sets}
\nomenclature[v, 29]{\( q_{i} \)}{ number of collocation nodes in ith dimension}
\nomenclature[v, 30]{\( r_{i}\)}{degree of the univariate gPC basis functions}
\nomenclature[v, 31]{\( \mathcal{S}\)}{crisp uncertainty set}
\nomenclature[v, 32]{\( s_{f}\)}{scaling factor}
\nomenclature[v, 33]{\( t\)}{time vector}
\nomenclature[v, 34]{\( t_{switch}\)}{ control switching time in the case study}
\nomenclature[v, 35]{\( \bm{u}\)}{open-loop control trajectory vector}
\nomenclature[v, 36]{\( \mathcal{U}_{a} \)}{admissible control}
\nomenclature[v, 37]{\( v\)}{objective function parameter in epigraph form}
\nomenclature[v, 38]{\( \tilde{\bm{x}}\)}{vector of uncertain basic quantities}
\nomenclature[v, 39]{\( \tilde{y}(\cdot)\)}{an arbitrary second-order variable or process}
\nomenclature[v, 40]{\( y_{p_{c}}\)}{ $PC$th-degree gPC approximation of $\tilde{y}(\cdot)$}
\nomenclature[v, 41]{\( \hat{y}_{m}(\cdot)\)}{ gPC expansion coefficients}

\nomenclature[v, 42]{\( \alpha_{w}\)}{quadrature weights}
\nomenclature[v, 43]{\( \Gamma \)}{finite domain of distribution function}
\nomenclature[v, 44]{\( \gamma \)}{normalization factor}
\nomenclature[v, 45]{\( \Delta t_{r} \)}{robust horizon}
\nomenclature[v, 46]{\( \eta\)}{size of crisp uncertainty set}
\nomenclature[v, 47]{\( \hat{\mu}_{\parm}\)}{unbiased estimate of the mean of $\tilde{\parm}$}
\nomenclature[v, 48]{\( \bm{\xi}\)}{state trajectories}
\nomenclature[v, 49]{\( \sigma_{\parm} \)}{ standard deviation of $\tilde{\parm}$}
\nomenclature[v, 50]{\( \mathcal{P}_{d}\)}{set of plants or polytope vertices or }
\nomenclature[v, 51]{\( \upsilon_{i}\)}{ $i$th vertex of the polytope}
\nomenclature[v, 52]{\( \phi_{k}(\tilde{x}_{i})\)}{univariate gPC basis functions}
\nomenclature[v, 53]{\( \Phi_{m}(\tilde{x}_{i})\)}{$n_{x}$-variate gPC basis functions}

\nomenclature[B, 1]{\( min\)}{ lower bound on $\parm$}
\nomenclature[B, 2]{\( max\)}{ upper bound on $\parm$}
\nomenclature[B, 3]{\( 0\)}{ initial}
\nomenclature[B, 4]{\( f\)}{ final}

\mbox{}

\printnomenclature

\end{document}